# Hellenic Open University Reconstruction & Simulation (HOURS) software package

User Guide & short reference of Event Generation, Cherenkov photon production and Optical Module simulation

A. G. Tsirigotis, G. Bourlis, A. Leisos and S. E. Tzamarias for the ASTRONEU collaboration

Physics Laboratory, School of Science & Technology, Hellenic Open University
Tsamadou 13-15 & Ag. Andreou, 26222 Patras, Greece
E-mail: tsirigotis@eap.gr

**Abstract**

In this document the simulation part of the Hellenic Open University Reconstruction & Simulation (HOURS) software package is described in detail. HOURS can be used for the generation, simulation, pattern recognition and reconstruction of high energy neutrino produced events in a very large volume neutrino telescope. The objective is to provide as accurate as possible a representation of event properties in a wide range of neutrino telescope configurations and medium optical properties. Moreover, HOURS contains software for the simulation and reconstruction of Extensive Air Showers (EAS) using the HEllenic LYceum Cosmic Observatories Network (HELYCON) scintillation counters. Using the information offered by the simulation/reconstruction of any EAS, and by considering the showers' energetic muons that penetrate the sea to the depth of the neutrino telescope, it is possible to study the joint performance of the neutrino and EAS detectors for physics or calibration purposes. HOURS has been used extensively for the optimization, development of calibration techniques and performance evaluation of the planned Mediterranean neutrino telescope, KM3NeT (km$^3$ Neutrino Telescope). The results of these studies have been published to various international scientific journals.

The code and further information may be found on the HOURS web page:
http://physicslab.eap.gr/EN/Simulation_software.html .

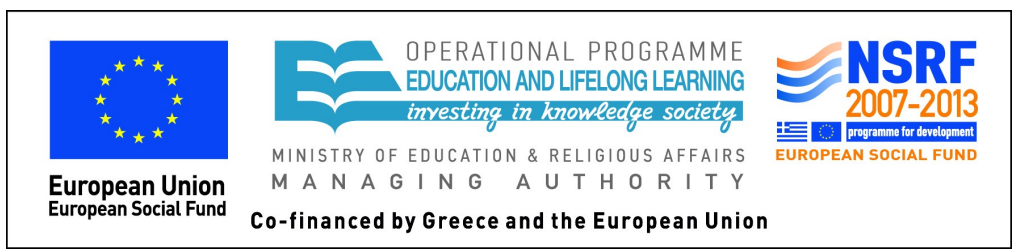

This research has been co-financed by the European Union (European Social Fund – ESF) and Greek national funds through the Operational Program "Education and Lifelong Learning" of the National Strategic Reference Framework (NSRF) - Research Funding Program: "THALIS - HOU - Development and Applications of Novel Instrumentation and Experimental Methods in Astroparticle Physics"

**Introduction**

The Hellenic Open University Reconstruction & Simulation (HOURS) physics analysis package for Astroparticle Physics contains physics event generators, event simulation, pattern recognition and filtering-triggering algorithms, and reconstruction analysis code. Some of these features are described in detail in the following chapters. HOURS has been used for the results of several publications and conference presentations regarding the accurate simulation and evaluation performance of the Hellenic Open University Extensive Air Shower detector (HELYCON) and underwater neutrino telescopes (NESTOR and KM3NeT) [1]. HOURS consists of two main simulation and reconstruction software chains. The first one is for the underwater neutrino telescope, while the second is for the Extensive Air Shower (EAS) surface detector arrays. This document describes the software developed for the first simulation chain regarding the tools used for the estimation of the performance of various techniques for the observation of astrophysical neutrino sources as well as several background rejection and calibration techniques. In Chapter 1 the software for neutrino and atmospheric shower event generation is described. In Chapter 2 the software for the detailed response of the underwater neutrino telescope to the generated neutrino interaction events or atmospheric muon/neutrino bundles is described. In Chapter 3 the photomultiplier and electronics simulation is described. In Fig. 1 the HOURS software chain flowchart diagram is presented, containing special routines for the detector element simulation and signal processing, pattern recognition methods as well as advanced tracking techniques. HOURS has been used for the performance estimation of a calibration technique for the KM3NeT, using the muons produced by energetic EAS.

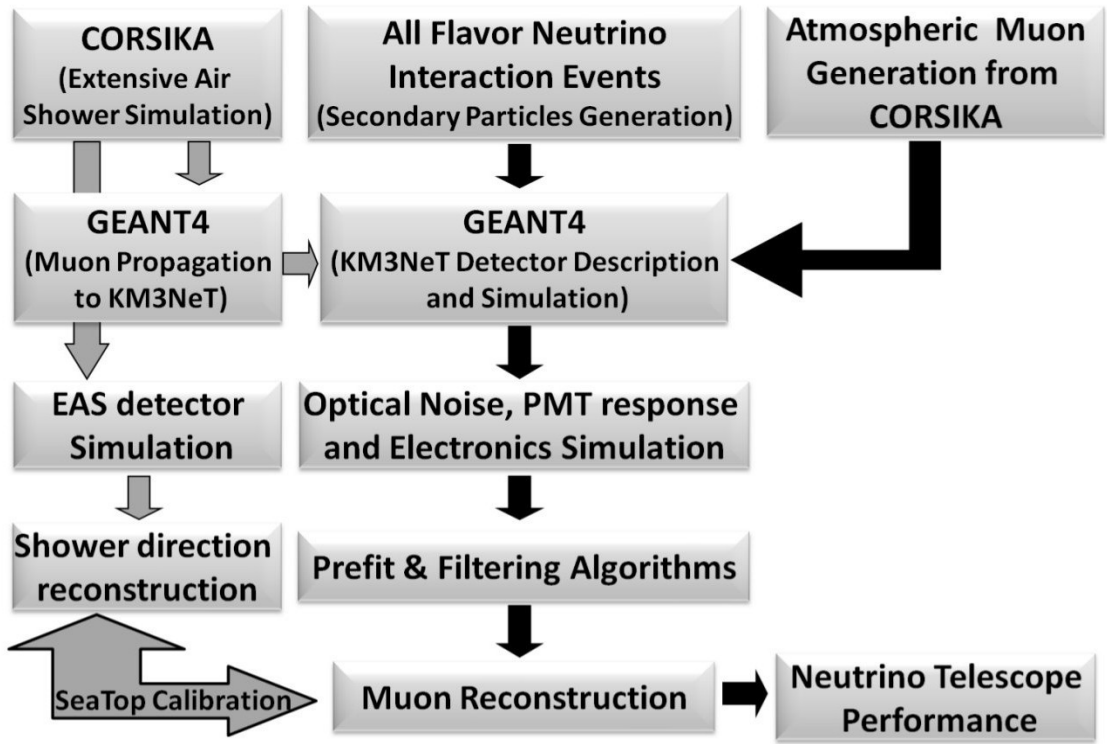

*Figure 1: The HOURS software chain flowchart diagram. The dark arrows indicate the simulation flow for the estimation of the detector response to neutrino interaction events, where the atmospheric muon background is generated by the EAS simulation package CORSIKA. The gray arrows indicate the simulation flow for the estimation of the performance of the Sea-Top calibration technique, where CORSIKA is used to produce the secondary particles of the EAS.*

# Chapter 1 – Very Large Volume neutrino Telescope (VLVnT) Event Generation

The first step of a detector simulation procedure is the generation of physics events. In the HOURS software package the primary neutrino momentum vector is generated according to several phenomenological models such as astrophysical sources and atmospheric neutrinos, while the shadowing of the Earth is taken into account. Although the Earth is transparent to neutrinos due to their weak interaction with matter, at higher energies above a few tens of TeV the neutrino interaction cross section becomes significant and neutrinos are shadowed by Earth's matter.

All flavor neutrino interactions with matter are described using the appropriate cross-sections, while the generation of secondary particles in these interactions is described by PYTHIA [2]. The atmospheric muon/neutrino background is described by using either Extensive Air Shower (EAS) simulators (CORSIKA [3]) to produce muons and neutrinos or by using atmospheric muon and neutrino parametrizations. In the first case the atmospheric muons are generated on the surface of the sea, while the produced atmospheric neutrinos are forced to interact in the vicinity of the detector. In the latter case both atmospheric muons and neutrinos are generated in the detector's neighborhood. More detailed information of the program structure can be found in the doxygen documentation.

The program comprises of about 6300 lines, is written in FORTRAN-77 and uses PYTHIA version 6.4.16. It also uses various routines from the general purpose library CERNLIB (version 2006). It also uses CORSIKA version 74xxx externally, of which the executable must be in the running directory. The program consists of 16 source code files:

1) HOURS_Gen_v8.1.f
   Main program containing various routines
2) atmoneutrino_agrawal.f
   Atmospheric neutrino flux parametrizations according to V. Agrawal et al
   http://arxiv.org/abs/hep-ph/9509423
3) atmoneutrino_honda.f
   Atmospheric neutrino flux parametrizations according to Honda et al
   http://arxiv.org/abs/astro-ph/0611418 , Phys.Rev.D75:043006,2007
4) atmoneutrino_bartol.f
   Atmospheric neutrino flux parametrizations (electron and muon (anti)neutrinos) according to bartol group - http://www-pnp.physics.ox.ac.uk/~barr/fluxfiles/
5) atmoneutrino_bartol_oscillated.f
   Atmospheric neutrino bartol fluxes after oscillation for normal or inverted hierarchy
6) atmoneutrino_conventional_semianalytic.f
   Semianalytic conventional atmoneutrino fluxes (only muon neutrinos) according to Gaisser - http://arxiv.org/abs/astro-ph/0104327 , Astroparticle Physics 16 (2002) 285-294
7) atmoneutrino_prompt.f
   Prompt atmospheric neutrino fluxes (all flavors, also tau) according to Rikard Enberg,Mary Hall Reno and Ina Sarcevic http://arxiv.org/abs/0806.0418 (Physical Review D78, 043005 2008)
8) grbneutrinos_waxman_bahcall.f
   Neutrino diffuse fluxes from Gamma Ray Bursts according to Waxman and Bahcall – arXiv:astro-ph/9701231 , arXiv:hep-ph/9807282
9) ajnjets_mannheim.f
   High energy neutrino fluxes from radio-loud AGNs according to Mannheim

(Astroparticle Physics, Volume 3, Issue 3, May 1995, Pages 295-302)
10) extrapgammasources_mannheim.f
    Diffuse neutrino fluxes from extragalactic proton-gamma sources according to Mannheim et al MPR - http://arxiv.org/abs/astro-ph/9812398
11) neutrino_td_sigl.f
    Neutrino fluxes from topological defects according to G. Sigl - http://arxiv.org/abs/astro-ph/9503014
12) gzkneutrinos_protheroe_johnson.f
    Neutrino fluxes due to GZK cutoff according to Protheroe & Johnson - http://arxiv.org/abs/astro-ph/9506119
13) gzkneutrinos_stanev.f
    Neutrino fluxes due to GZK cutoff according to Stanev - Nuclear Physics B (Proc. Suppl.) 136,2004,103-110
14) divdif.f
    One dimensional polynomial interpolation routine adopted from CERNLIB and changed to use double precision reals
15) pop4_1303_3565.f
    The input primary spectrum used for reweighting CORSIKA events. It is based on global fit with population 4 of http://arxiv.org/abs/1303.3565
16) HOURS_Gen.inc
    File included in subroutines and containing commonly used variables

In addition there is a number of data files distributed with the program that contain the CTEQ10 cross sections and bartol fluxes with or without taking into account neutrino oscillations.

The program has been tested to work with compilers g77 and gfortran on Linux fedora core 4-17 operating systems and gcc versions 3.4.x , 4.0.x , 4.1.x , 4.3.x , 4.4.x, 4.7.x . The compilation script is included in the distribution, as well as various examples.

**1.1 Program running and input**

For neutrino event generation and CORSIKA runs the program runs using the following command:

./HOURS_Gen_v8.1 run_number seed1 seed2 seed3 seed4 output_file  < INPUT_file where the first two seeds are for the initialization of the random generators of CERNLIB and PYTHIA respectively. The third and fourth seeds are for the initialization of CORSIKA random number generators.

In the case of neutrino event generation, the output file describes the kinematic parameters and types of neutrinos that interact in the vicinity of the detector, as well as the secondaries of the interaction. In the case of CORSIKA runs for the generation of events emanating from atmospheric showers, the output file describes the kinematic parameters of the atmospheric muons capable to reach the detector's instrumented volume and/or atmospheric neutrinos created by the same shower. In both cases event weights are included in the output file. The output file is in text format which is described in Appendix A.

Running examples of the program can be found in the distribution. The two of them are neutrino interaction generation examples, for muon and electron neutrinos respectively. The other two are CORSIKA atmospheric muon and neutrino event generation examples. In one of these examples, atmospheric neutrino interactions are also included (see 20th input parameter below).

The input file is a text file containing various parameters steering the event generation (the parameter values can be typed on the screen if no input file is specified). The sequence of the input variables in the input file is the following (typing ./HOURS_Gen_v8.1 help the program gives a short description of command line arguments and the format of the input file):
1) Simulation mode (int): see subsection below
2) Differential spectral index assuming a power law energy spectrum (negative real).
3) Neutrino type or CORSIKA primary (int):
   1,2: electron neutrino and antineutrino, respectively
   3,4: muon neutrino and antineutrino, respectively
   5,6: tau neutrino and antineutrino, respectively
   For Extensive Air Showers the particle code is as described in CORSIKA documentation
4) Flag for selection of medium targets in neutrino interactions (int):
   0: Only nucleon targets
   1: Nucleon and electron targets
5) Flag for Earth's shadowing (int):
   0: shadowing due to Charged Current (CC) neutrino interactions only
   1: shadowing due to CC & Neutral Current (NC) neutrino interactions
6) Flag for generation of neutrino interactions around detector (int):
   1: only contained interactions in the CAN. The CAN is defined as a cylindrical volume enclosing the detector. The bottom of the CAN is the sea floor. The top of the can is a number of maximum optical absorption lengths above the upper optical modules of the detector (see parameters 17,18 below). The radius of the CAN is the same number of maximum optical absorption lengths larger than the radius of the detector's instrumented volume.
   0: also non contained neutrino interactions
7) Flag for generation mode of atmospheric neutrinos with neutrino oscillations (mode 14)(int):
   1: normal neutrino mass hierarchy
   -1: inverse neutrino mass hierarchy
8) Minimum energy of neutrinos or CORSIKA primaries in GeV (real)
9) Maximum energy of neutrinos or CORSIKA primaries in GeV (real)
10) Minimum zenith angle (degrees – real – 0 degrees correspond to vertical downgoing incident neutrinos or CORSIKA primaries)
11) Maximum zenith angle (degrees – real)
12) Number of events to generate (int)
13) Depth of the detector's center (cm – positive real)
14) Position of the sea floor (cm – negative real – where zero is at the detector center and z-axis points upwards)
15) Half height of the detector's cylindrical instrumented volume (real in cm).
16) Horizontal radius of the detector's cylindrical instrumented volume (real in cm)
17) Maximum optical absorption length (real in cm)
18) Number of maximum absorption lengths to extend the detector's instrumented volume and define the CAN (real)
19) 
   - Lower cut of neutrino's energy in CORSIKA runs when the generation also of the atmospheric neutrino interactions are required (real in GeV)
   - Energy factor of neutrino exponential energy cut in neutrino interaction runs (real in GeV). If this is zero then the neutrino energy spectrum is a power law spectrum for all energies. If it is positive the spectrum is $\sim E^{-n}e^{-E/E_c}$. If it is negative the spectrum is $\sim E^{-n}e^{-\sqrt{-E/E_c}}$.
20) Flag to produce atmospheric neutrino interactions from CORSIKA runs (int):
   0: only atmospheric muons

  1: atmospheric neutrino interactions also
    From the atmospheric neutrino bundle one neutrino is chosen randomly to interact in the vicinity of the detector
21) Number of time to repeat each CORSIKA event, each time shifting the shower axis (int)
22) File name of the CORSIKA executable (without path - must be in execution directory)

**1.1.1 Simulation modes**

The simulation modes (first argument in the input parameters) applicable are:
0: Atmospheric muon (and neutrino) bundles from CORSIKA,
1: Atmospheric neutrino flux generation according to AGRAWAL ET AL, arXiv:hep-ph/9509423, valid for electron neutrino and antineutrino, muon neutrino and antineutrino and energies from 1 GeV to 10 TeV
2: Neutrino flux generation from topological defects, G. Sigl, arXiv:astro-ph/9503014 , valid for electron neutrino and antineutrino, muon neutrino and antineutrino and energies from 10 TeV to 100 EeV
3: Neutrino flux generation produced from protons near GZK energies interacting with the Cosmic Microwave Background (CMB), valid for electron neutrino and antineutrino, muon neutrino and antineutrino and energies from 200 PeV to 1 ZeV
4: Neutrino flux generation from extragalactic proton acceleration sources, valid for muon neutrino and antineutrino and energies from 1 TeV to 100 EeV
5: Neutrino flux generation from AGN Jets, valid for muon neutrino and antineutrino and energies from 100 TeV to 100 EeV.
6: Neutrino flux generation from Gamma Ray Bursts (GRBs), arXiv:astro-ph/9701231 , arXiv:hep-ph/9807282, valid for muon neutrino and antineutrino and all energies.
7: Neutrino flux generation produced from protons near GZK energies interacting with CMB (same physics different model from the one of mode 3), valid for electron neutrino and antineutrino, muon neutrino and antineutrino and energies from 1 TeV to 1 ZeV
8: Neutrino flux generation with a $E^a$ energy spectrum
9: Conventional atmospheric neutrino flux generation according to HONDA ET AL, arXiv:astro-ph/0611418, valid for electron neutrino and antineutrino, muon neutrino and antineutrino and energies from 0.1 GeV to 10 TeV
10: Conventional atmospheric neutrino semianalytic flux generation according to GAISSER, arXiv:astro-ph/0104327, calculated from 10 TeV to 100 PeV and renormalized so that to coincide with HONDA ET AL, arXiv:astro-ph/0611418 flux at 10 TeV. Valid for muon neutrino and antineutrino and energies from 10 TeV to 100 PeV
11: Prompt atmospheric neutrino flux generation according to Rikard Enberg, Mary Hall Reno and Ina Sarcevic, arXiv:0806.0418, valid for all neutrino flavors (tau neutrino also) and energies from 100 GeV to 100 PeV.
12: Atmospheric neutrino flux generation including all components 9-11 above, valid for muon neutrino and antineutrino and energies from 0.1GeV to 100 PeV.
13: Generation with the estimated spectrum of the candidate neutrino source RXJ1713.7 according to S.R. Kelner, F.A. Aharonian and V.V. Bugayev, arXiv:astro-ph/0606058
14: For generation with oscillated atmospheric neutrino bartol spectrum, where initial spectrum is the one for following mode normal and inverted hierarchy with the mass differences and oscillation angles are from middle 2012 latest data .
15: For generation with the bartol spectrum
 http://www-pnp.physics.ox.ac.uk/~barr/fluxfiles/0408i/index.html
 for high energy (10GeV-10TeV) and

http://www-pnp.physics.ox.ac.uk/~barr/fluxfiles/0403i/index.html
for low energy (0.1GeV-10GeV) Kamioka solar minimum zenith distributions with 20 bins per decade in logE .

### 1.1.2 Decay of short lived particles

Very short lived particles produced by PYTHIA in neutrino interactions are forced to decay since they are not included in the GEANT4 framework or their decay channels are not defined. For GEANT-4.9.6 that is used in HOURS these particles have PDG (Particle Data Group) codes:
- 411 (Charmed meson $D^+$),
- 421 (Charmed meson $D^0$),
- 431 (Charmed-strange meson $D_s^+$),
- 511 (Bottom meson $B^0$),
- 521 (Bottom meson $B^+$),
- 531 (Bottom-strange meson $B_s^0$),
- 4122 (Charmed baryon $\Lambda_c^+$),
- 4132 (Charmed baryon $\Xi_c^0$),
- 4232 (Charmed baryon $\Xi_c^+$),
- 4332 (Charmed baryon $\Omega_c^0$),
- 5122 (Bottom baryon $\Lambda_b^0$),
- 5132 (Bottom baryon $\Xi_b^-$),
- 5232 (Bottom baryon $\Xi_b^0$)
- and their antiparticles.

### 1.2 Atmospheric muon and neutrino bundle generation

Atmospheric muon background are by far the most abundant background in a neutrino telescope, the event rate being tens of millions events per day for a detector with a few cubic kilometers of instrumented volume. In comparison the expected atmospheric neutrino rate interacting in the vicinity of the detector and reconstructed is a few hundred per day, while the most luminous astrophysical source is expected to give a few detectable events per year. The fact that the atmospheric muon background rate is many orders of magnitude more than any neutrino signal rate presents a difficulty. The need of a comparable to the detector experimental time atmospheric muon sample lifetime results in a large amount of simulated atmospheric shower events, of the order of hundreds of billions. A way to deal with the large amount of shower events needed and the fact that atmospheric shower simulation is CPU time consuming, is the application of parametrizations of the kinematical characteristics of the atmospheric muon bundles reaching the detector's depth (see for example the package MUPAGE [4]). This method has the advantage of relatively quick atmospheric muon generation, but the disadvantage that the parametrizations must be redone for every model of the primary composition used. In the event generation package of HOURS for the generation of atmospheric muon background events we use atmospheric shower events from CORSIKA without any parametrization of the atmospheric muon flux.

CORSIKA is used to produce the secondary particles of the Extensive Air Showers (EAS) with simulation of the events according to a chosen spectral index and primary particle type. Then the events are reweighed according to the model described in reference [5]. However all the information needed to reweigh the events according to any model of primary composition and flux is written in the output. So the user could reweigh the events according to any preferred model of primary cosmic rays. For the simulation any executable of the CORSIKA program can be used, with the restriction that the output format should be the one produced when THINning is enabled during compilation, while COMPACT and HISTORY options are not enabled. The executable should be also for volume detectors (VOLUMEDET option during CORSIKA compilation).

The shower information kept in these simulations are the primary particle type, energy, zenith and azimuth angle, as well as the kinematical parameters of all the muons that are capable of penetrating the sea and reach the detector's instrumented volume.

With the generation of atmospheric muon bundles with CORSIKA there is the possibility to generate also the secondaries of the interactions of the accompanying atmospheric neutrinos of the same shower. This operation mode, which is enabled using the 20th parameter of the input file, has been included in order to study the performance of the detector to detect ultra high energy cosmic neutrinos that are coming from above the horizon. Upcoming ultra high energy neutrinos with energies above 100TeV are shadowed by the Earth and their detected flux is decreased. Fig. 2 shows the earth shadowing effect for high energy neutrinos, while Fig. 3 presents the Earth's matter profile for which this shadowing is estimated. The only alternative to increase the detected flux is to detect nearly horizontally incident neutrinos or downcoming ones. However, the atmospheric neutrino flux near the horizon is increased, so there is more background when detecting ultra high energy cosmic neutrinos incident horizontally. On the other hand, there is the possibility to distinguish cosmic neutrinos from atmospheric ones by using the fact that downcoming atmospheric neutrinos at such high energies are almost always accompanied by atmospheric muons energetic enough to reach the detector depths.

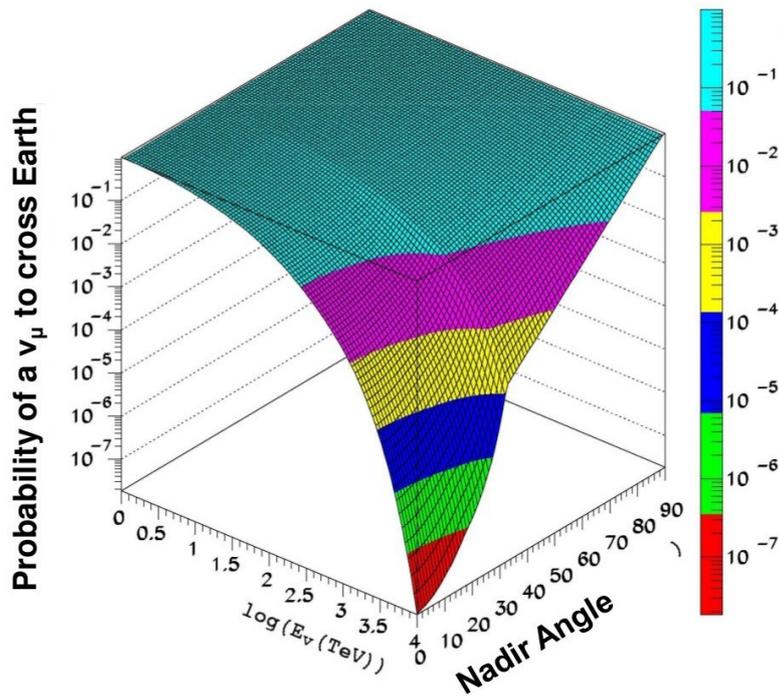

*Figure 2: The probability a neutrino to cross the Earth and reach the detector's neighborhood as a function of the neutrino's energy and nadir angle. For nadir angles less than 70 degrees and energies more that 100TeV this probability is less than 10%.*

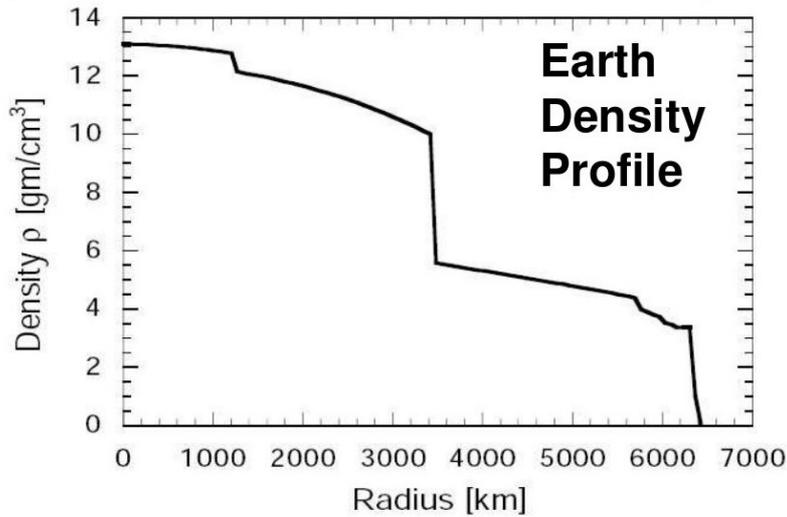

*Figure 3: The density of the Earth as a function of the distance from the center, according to the Preliminary Reference Earth Model (PREM).*

### 1.3 Results from atmospheric muon and neutrino generation

In the following results produced using the event generation software of HOURS are presented. Many of the results are produced using also the HOURS_KM3Sim program explained below, which except for the description of detector performance can transport shower particles to the detector neighborhood.

### 1.3.1 Atmospheric muon bundle multiplicity and event rate

The majority of atmospheric muon events are not consisting of single muons especially those events emanating from high energy extensive air showers. The muons come in bundles with muon multiplicity counting up to many hundreds of muons per event. The spatial and temporal extension of the high energy muons when they reach the detector are of the order of a few meters and a few tens of nanosecond respectively. In Fig. 4 the multiplicity distribution and the event rate of atmospheric muon bundle is presented. The shower events have been simulated by CORSIKA using the primary composition model of reference [6]. The muons from the shower are registered at the sea surface and HOURS_KM3Sim has been used for the transportation to the detector deployed at a sea depth of 3500m.

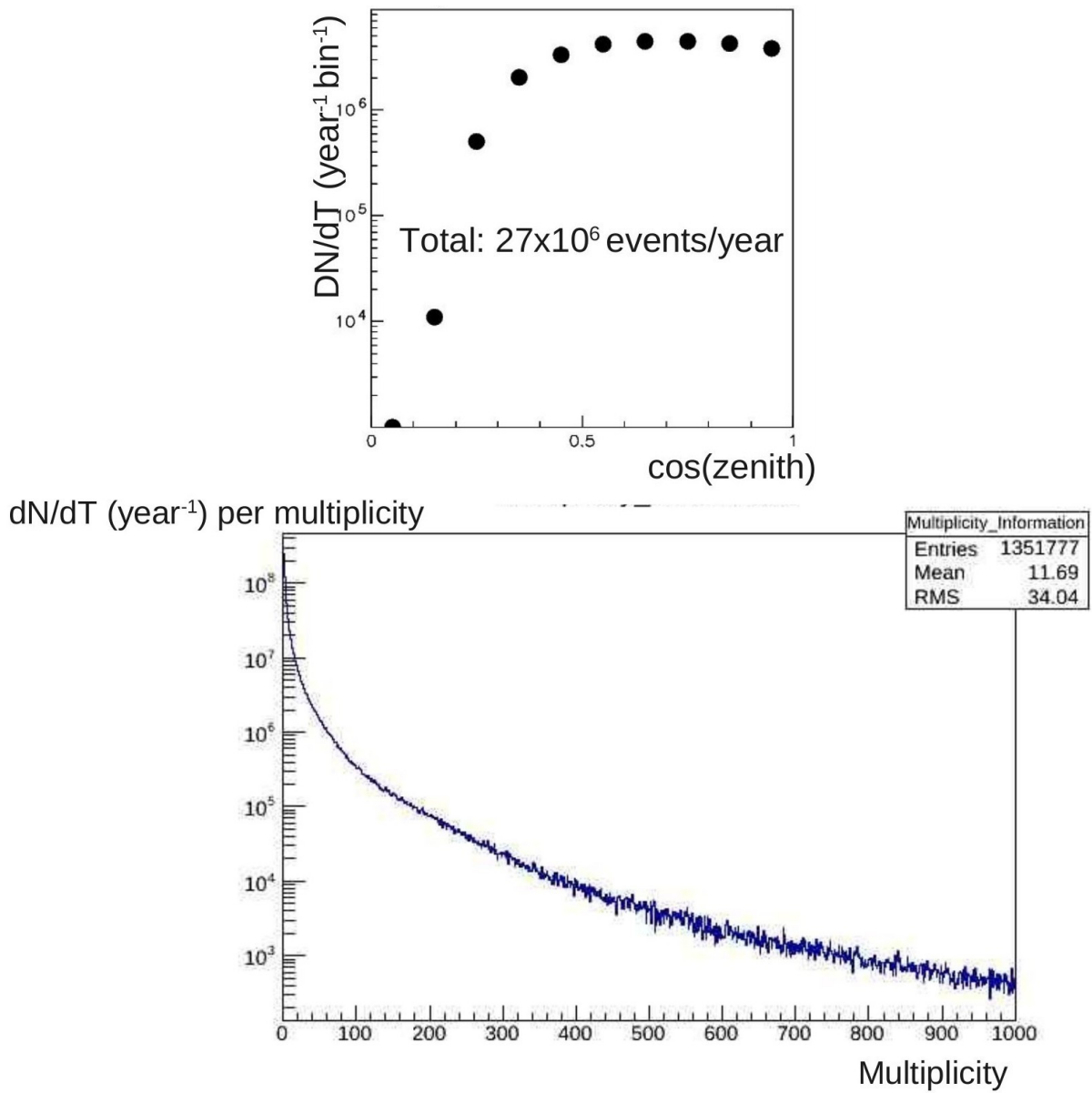

*Figure 4: The event rate (upper plot) of atmospheric muon bundles for a neutrino telescope deployed at a depth of 3500 m and with an instrumented volume 0.5 km3. The events shown are for muon bundles with total energy grater than 100GeV when entering the detector's instrumented volume. The lower plot shows the distribution of the number of muons per bundle (multiplicity). As seen from the lower plot, the multiplicity extends up to thousands of muons per bundle.*

### 1.3.2 Atmospheric neutrino event rate

The atmospheric neutrinos represent an irreducible background for neutrino telescopes aiming at detecting cosmic neutrino fluxes. This is especially true when detecting upcoming neutrinos since there is no additional information that can be used to distinguish upcoming atmospheric neutrinos from upcoming cosmic neutrinos. There is no way to distinguish these two types of neutrinos on an event by event basis and the only way to detect a cosmic signal from sources below the horizon is when the astrophysical source is a point source and the number of events from the direction of the source exceeds significantly the number of the expected events from the background of atmospheric neutrinos. However, high energy downcoming atmospheric neutrinos can be distinguished from cosmic ones, as explained in the next Section. Fig. 5 shows the atmospheric neutrino flux as a function of energy and zenith angle, while in Fig. 6 a comparison is presented between neutrino flux from HOURS_Gen event generator and semi-analytical calculations based on the same primary composition model (from reference [7]).

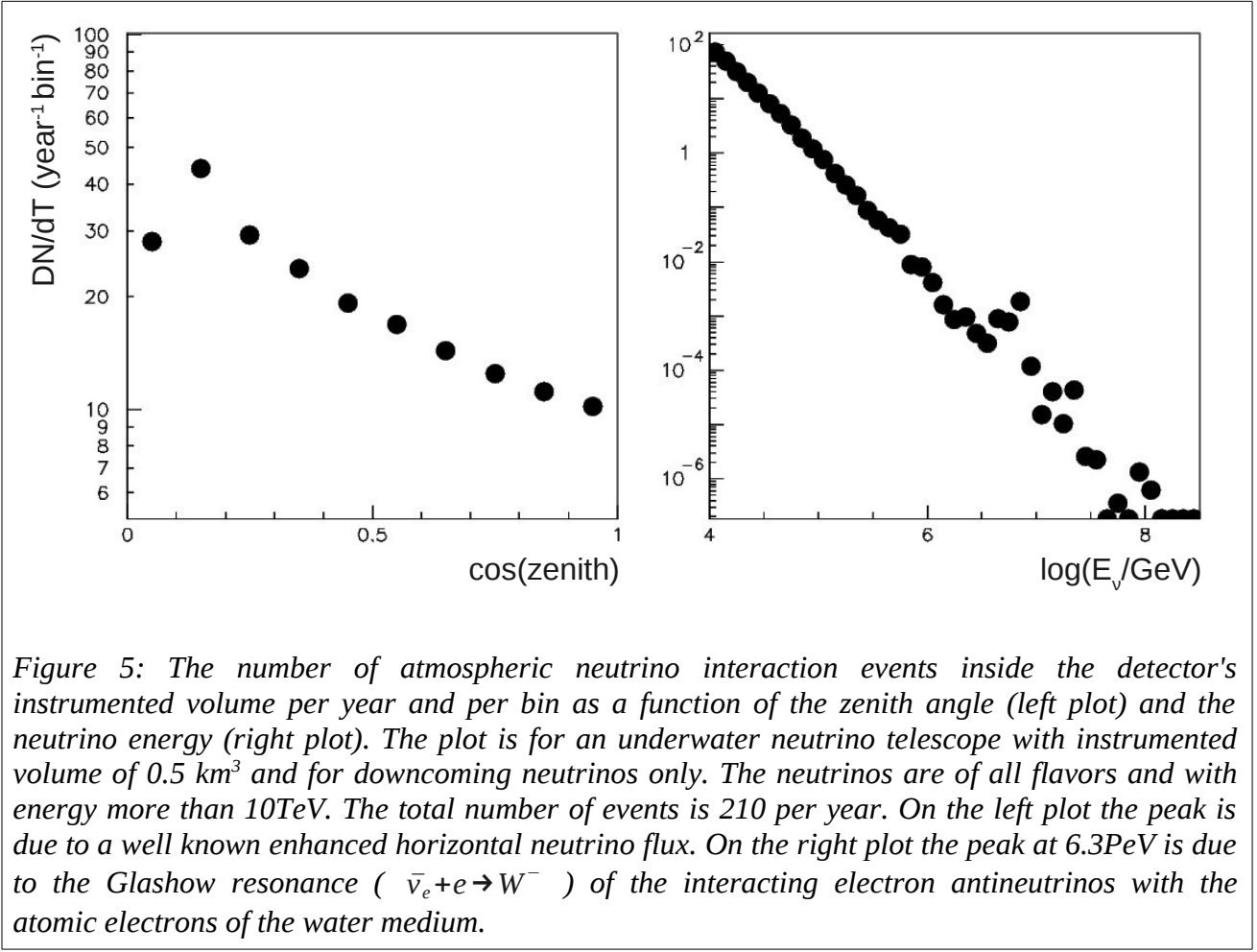

*Figure 5: The number of atmospheric neutrino interaction events inside the detector's instrumented volume per year and per bin as a function of the zenith angle (left plot) and the neutrino energy (right plot). The plot is for an underwater neutrino telescope with instrumented volume of 0.5 $km^3$ and for downcoming neutrinos only. The neutrinos are of all flavors and with energy more than 10TeV. The total number of events is 210 per year. On the left plot the peak is due to a well known enhanced horizontal neutrino flux. On the right plot the peak at 6.3PeV is due to the Glashow resonance ( $\bar{v}_e + e \rightarrow W^-$ ) of the interacting electron antineutrinos with the atomic electrons of the water medium.*

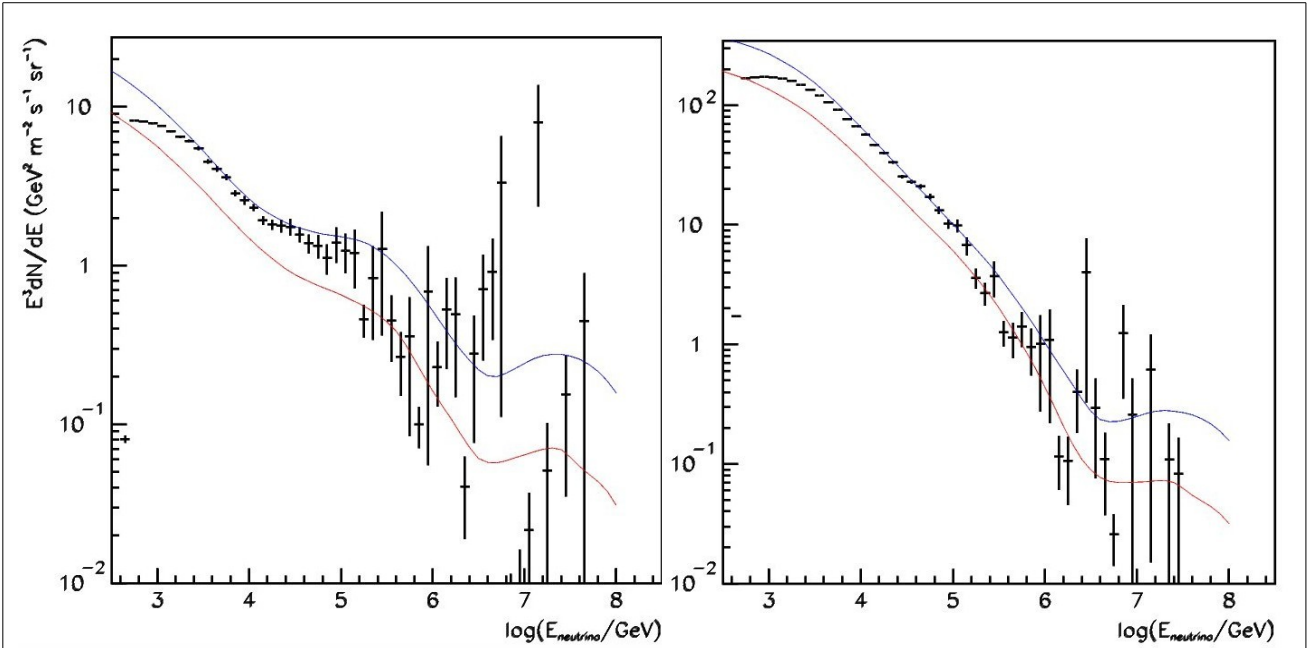

*Figure 6: Vertical atmospheric neutrino flux (points) as produced from HOURS_Gen compared to semi-analytical calculations according to the GST 3-gen model of reference [2] and SYBILL (blue line) or QGSJET01 (red line) high energy hadronic interaction models. Left plot is for electron neutrinos and right plot for muon neutrinos. The produced spectrum from HOURS_Gen levels off below a few TeV because the primary cosmic rays are generated with energies more than 10TeV.*

**1.3.3 Atmospheric neutrino event rate with accompanying atmospheric muons**

As explained in the previous Section upcoming atmospheric neutrinos cannot be distinguished by cosmic ones. However, high energy downcoming neutrinos (>10TeV) are created by energetic showers that also produce atmospheric muons capable to penetrate the sea and reach the detector. The signal from these muons can be used to distinguish high energy downcoming atmospheric neutrinos from cosmic ones. The method is simple and it has already been used by ICECUBE to enhance the detection of high energy cosmic neutrinos, resulting to the first extraterrestrial high energy neutrino flux observed by the South pole neutrino telescope.

The downcoming events in a neutrino telescope are generally classified in two types, according to their topology:
1. Events such that the first hits are at the upper or side edges of the detector. These events could be:
    a) Downcoming atmospheric muons (top picture of Fig. 7).
    b) Downcoming atmospheric muons accompanied by an atmospheric neutrino produced by the same EAS and interacting inside the detector (middle picture of Fig. 7).
    c) Downcoming cosmic neutrinos interacting outside the detector (top picture of Fig. 7).
2. Events such that the first hits are well inside the detector. These events could be:
    a) Downcoming atmospheric muons that sneak through the outer layers of the detector and suffer a catastrophic radiative process inside the detector (bottom picture of Fig. 7).
    b) Downcoming atmospheric neutrinos interacting inside the detector without having detectable accompanying atmospheric muons produced by the same EAS (bottom picture of Fig. 7).
    c) Downcoming cosmic neutrinos interacting inside the detector (bottom picture of Fig. 7).

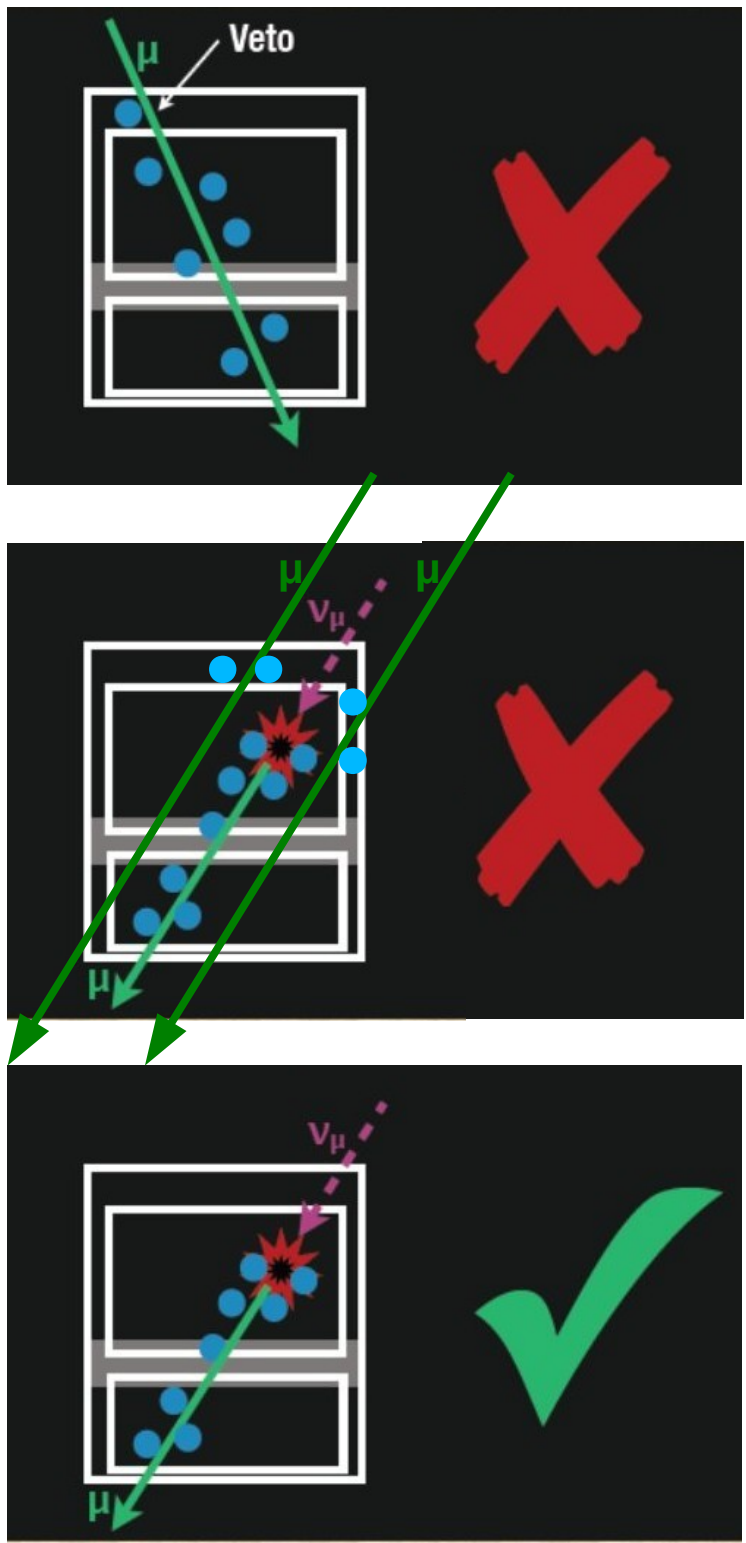

*Figure 7: Illustration of three types of events described in the text. Top picture describes muons created outside the detector activating first the upper layer of the detector. Middle picture describes atmospheric events with muons activating the upper layer of the detector and a neutrino interacting inside. Bottom picture describes the most interesting events for the estimation of cosmic neutrino spectrum. In these types of events the secondaries of a neutrino interaction activate the inner optical modules of the detector, while the outer ones have not a coincident signal.*

The case 2.c is of interest for the discovery of a cosmic neutrino flux. The events described by classification type 1.c cannot be distinguished from 1.a, and moreover in the case 1.c the neutrino energy cannot be estimated. The signal events classified in case 2.c have as background events classified in cases 2.a and 2.b. However, by asking for high energy events of case 2.c by making a lower cut on the number of detected photons, we can also exclude a large percentage of events 2.a and 2.b. This is because the majority of high energy atmospheric events except of the interacting atmospheric neutrinos are also producing muon bundles incident on the detector. The spatial and temporal distribution of these muons relative to the interacting neutrino is of the order of a few tens of meters and nanoseconds respectively, as is presented in Fig. 8. In these plots only events with the interacting neutrino's energy larger that 10 TeV are included. The decrease of the distance as a function of muon's energy in the top plot is because of the decrease of the muon's angular deviation due to multiple scattering and radiative processes. The lower plot shows no such behavior since the multiple scattering and radiative processes do not affect the arrival time of the muons. The approximately constant arrival time differences in the lower plot is a feature derived from the extensive air shower's core.

Another question in mind when trying to exclude the background of downcoming atmospheric neutrinos and detect a diffuse flux of cosmic neutrinos is how often the interacting atmospheric neutrinos in the detector are accompanied by atmospheric muons that trigger the outer layers of the detector. The answer can be seen in the plot of Fig. 9. As is evident from the plot the percentage of events containing a muon bundle incident on the detector and coming along with the interacting neutrino depends on both the neutrino's energy and zenith angle. This is expected as more energetic neutrinos are produced from more energetic showers that also produce more energetic muon bundles capable to penetrate the sea and reach the detector's depth. Also increasing the zenith angle the muons have to cross an increased slant depth and therefore they loose more energy with some of them absorbed by the sea water. So the rejection efficiency of the atmospheric neutrino background depends both on neutrino energy and zenith angle. As it can be seen in Fig. 9, 90% of the interacting atmospheric neutrinos with energy greater than 40 TeV are accompanied by a muon bundle of energy at least 100GeV and for zenith angles up to 60 degrees. Of course the rejection efficiency depends also from the performance of the neutrino telescope to detect the atmospheric muon bundle that trigger the outer layers of the telescope.

Taking into account the detector performance to detect the atmospheric muon bundle and including all detector component simulation the result of the atmospheric neutrino detection efficiency is presented in Fig. 10. The left plot is the downcoming atmospheric neutrino rate for all flavors as a function of the neutrino's energy (same as Fig. 5 right), while the right plot is the rate that remains after the application of a veto technique in a 0.5 km$^3$ neutrino telescope. The telescope consists of 115 vertical strings of optical modules, with each string carrying 18 of them. The detector has a roughly cylindrical shape with diameter about 1 km and height 612 m while the strings are arranged homogeneously. Several veto setups have been tested and the best one found defines the outermost layer of the strings as well as the 3 top layers of the detector as veto area. The veto technique consists of recognizing and rejecting events that are most likely to be atmospheric neutrinos interacting inside the detector. The events are recognized as such when the veto area contains more than a predefined number of early activated photomultipliers (PMTs). Moreover in order to get rid of low energy neutrino interaction events the inner area must have enough activated PMTs. The events used are from simulation of extensive air showers by the event generator HOURS_Gen using CORSIKA. The showers are simulated isotropically with zenith angles up to 87 degrees and energies from 10 TeV to 10 EeV. As it can be seen with comparison of the two plots of Fig. 10, the rejection efficiency for events with energy more than 40 TeV is about 90%. This the same as the efficiency concluded from Fig. 9. The events with energy less than 40 TeV are strongly suppressed due to the cut on the number of activated PMTs in the inner detector area.

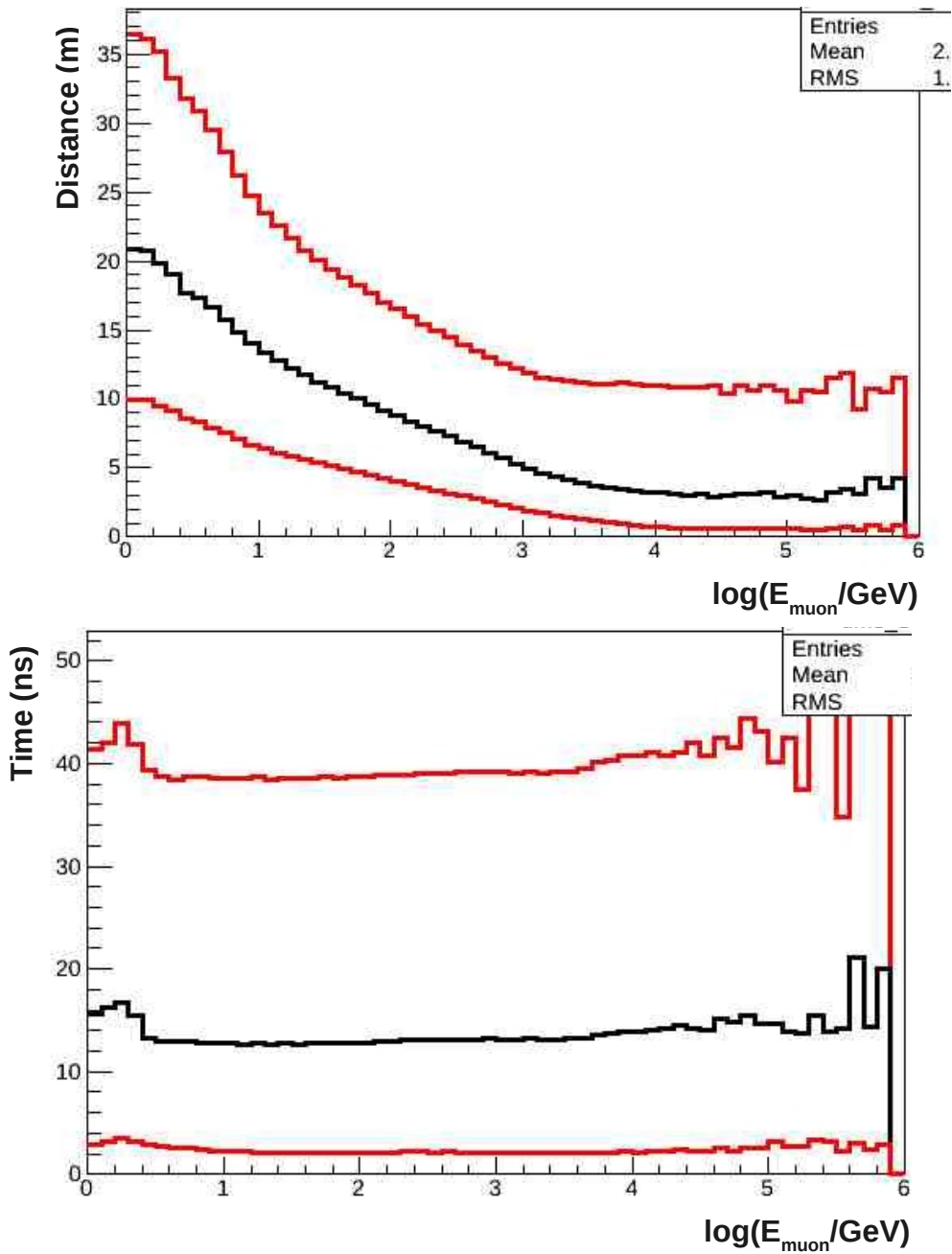

*Figure 8: The spatial (top) and temporal (bottom) profile of the muons with respect to the atmospheric neutrino of the same shower interacting inside the detector. The muon track distances from the neutrino track and the respective arrival time differences are shown as a function of the energy of the muon when it is entering the detector. The black histograms are the median of the distance and arrival time difference distributions for each energy bin, while the red histograms are the 16% and 84% quantiles of the same distributions.*

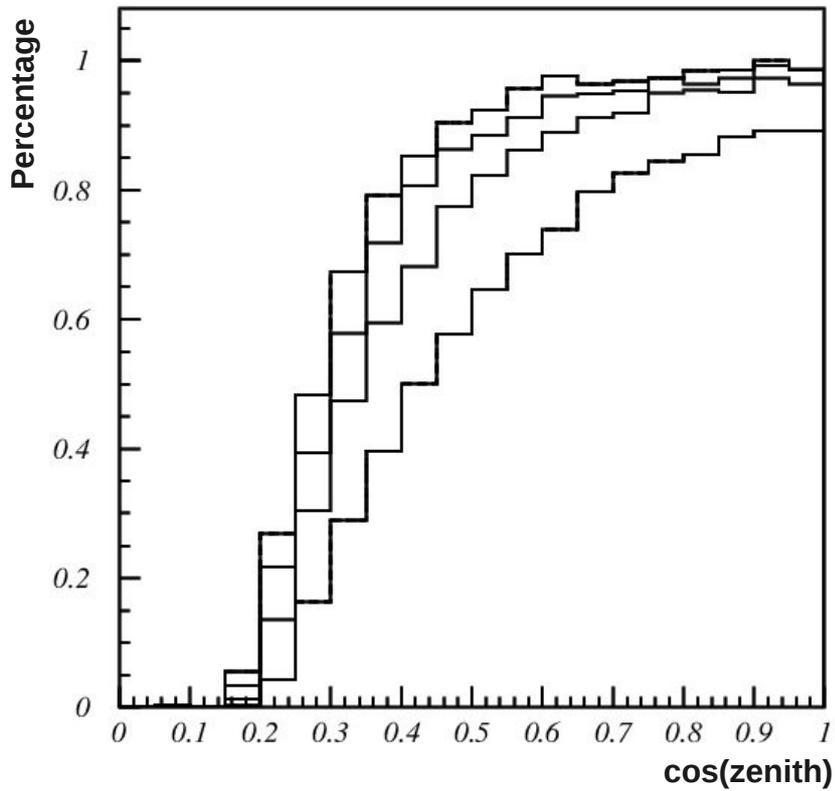

*Figure 9: The percentage of atmospheric neutrinos interacting inside the detector that are accompanied by an atmospheric muon bundle with energy greater than 100 GeV. The different histograms are for different cuts on neutrino energies. From bottom to top the neutrino energies are greater than 10, 20, 30, 40 TeV. As it can be seen more than 90% of neutrinos with energy greater than 40 TeV have muon bundles for zenith angles less than 60 degrees (cosine 0.5).*

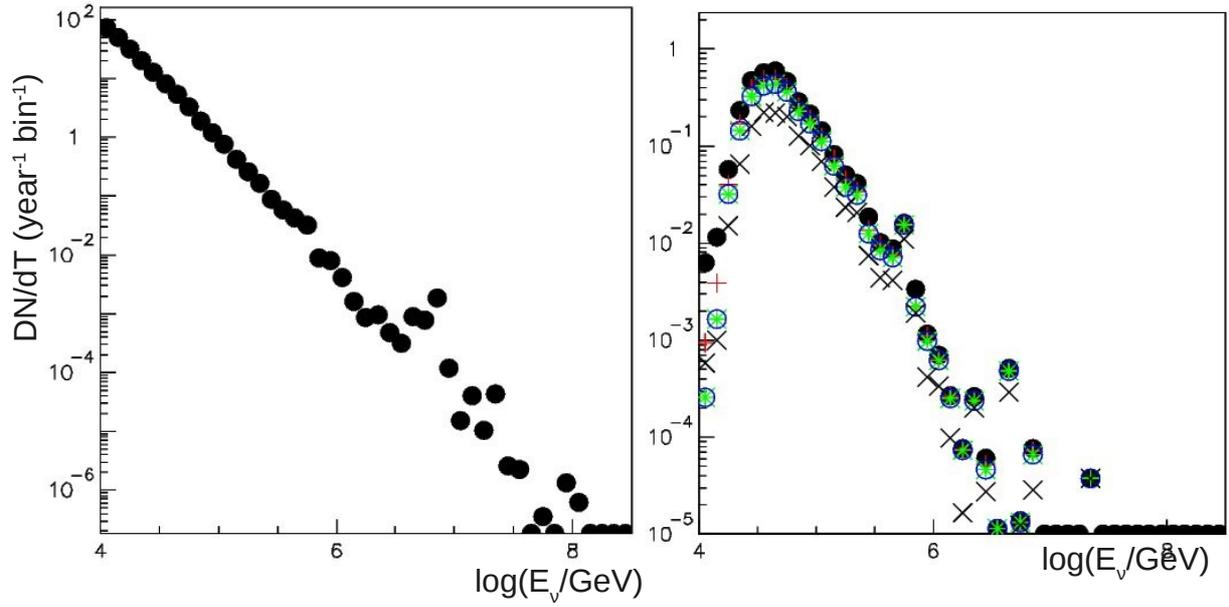

*Figure 10: The rate of the downcoming atmospheric neutrino interaction events in a detector of 0.5 km$^3$ instrumented volume (left plot). The total number of events per year is 210. The right plot is the rate of the same events after cuts applied that use the veto technique. The different points on the right plot are for different veto setups. The veto setup described in the text corresponds to the green stars. For this setup the total number of events per year remaining is 2.42.*

# Chapter 2 – Detector description and Cherenkov photon production.

The detector description part of HOURS – HOURS_KM3Sim – describes in details the detector architecture and all the physical processes resulting to experimental signatures in the underwater neutrino telescope. The detector is assumed to be constituted of photomultiplier tubes (PMTs) enclosed in spherical glass pressure housing, which are optically coupled with the photo cathode. The whole construction is called Optical Module (OM) and is the basic unit of a neutrino telescope. When a muon (or other charged particle) is moving faster than light in the medium (sea water) around the detector Cherenkov light is emitted and it is collected by the PMTs. Using the PMT's signal characteristics the kinematical parameters of the particles emitting these photons can be estimated.

The particles input in the detector description software are the ones produced by the generation part of HOURS, as described in Chapter 1. These particles are either secondaries of neutrino interactions in the vicinity of the detector or are atmospheric muons emanating from extensive air showers and have enough energy to penetrate the sea and reach the top of the detector.

In the detector description part of HOURS the simulation procedure is factorized in the following main stages:
- The simulation of the interaction of the input particles and their secondaries in the water surrounding the detector.
- The simulation of the production of Cherenkov photons and their propagation to the PMT's photocathodes.

## 2.1 Simulation of interactions in the Water – Light Emission and propagation

In this subsection we describe the simulation of all the possible particle interactions that are taking place in the water. This simulation stage is implemented in an autonomous software package and is build upon the GEANT4 general simulation kit [8]. The additional code is written in C++. The package contains about 12500 lines of source code. The most important classes of the program are (for more details refer to the doxygen documentation):
- KM3Physics
- KM3Detector
- KM3Cherenkov
- KM3SD
- KM3PrimaryGeneratorAction
- KM3EMShowerModel
- KM3EMDirectFlux
- KM3HAShowerModel
- KM3EMDeltaFlux
- KM3StackingAction
- KM3MuonParam

In the following we present the functionality of these classes and some characteristic plots showing the results of the simulation.

### 2.1.1 Class KM3Physics

In this class we define the particles and processes, which must be included in the simulation. Not all of the possible processes included in GEANT4 toolkit are used, but these that could give a detectable signature in the event's signal. The processes included in the simulation are presented in Table 1.

| Particle | Process |
|---|---|
| Charged | Multiple scattering, ionization, Cherenkov |
| Unstable | Decay, radioactive decay (for unstable ions) |
| μ+, μ- | Bremsstrahlung, muon nuclear interaction, e+e- pair production |
| μ- | Capture at rest |
| e+, e- | Bremsstrahlung, Electromagnetic (EM) shower fast simulation |
| e+ | annihilation |
| γ | Conversion to e+/e- pair, Compton scattering, photoelectric effect, EM shower fast simulation |
| Optical photon | Absorption, Rayleigh and Mie (particulate) scattering |
| Hadrons, Ions | Hadronic elastic and inelastic interaction processes |
| π-, k- | Capture at rest |
| π+,π-,k+,k-, $k_L^0, p, \bar{p}, n, \bar{n}$ | Hadronic shower fast simulation |

*Table 1: The processes included in the simulation*

Also the range cuts applied for the simulation of particles are defined in this class. For high energy events this range cut is defined to be 0.5mm, which corresponds to the electron cherenkov limit (Ekin=240 keV). For low energy events (interaction events of neutrinos from supernovae, or $^{40}$K decay events for background simulation) the range cut is set to a vary small value of 1μm, in order to account for the produced electron/positron multiple scattering and produce the correct hit correlations between the photomultipliers of the same optical module.

The range cut in GEANT4 is a cut on the range of the secondaries produced by the particle this range cut is applied. E.g. for a range cut of 0.5 mm on a muon, no atomic electrons from muon ionization are produced if their range is less than 0.5 mm in water. But although these ionization electrons are not produced, the energy loss of the muon due to this process is taken into account. The range cut should be such that no particles capable of emitting Cherenkov radiation are lost, and also the recoil of their parent particle can be neglected. When detecting high energy events (multi GeV events) the range cut should be such that the electrons from ionization above the Cherenkov threshold are emitted, while the recoil of their parent particles for lower energy electrons is negligible. When detecting low energy events (sub GeV events, such as those from $^{40}$K decays or from interactions of supernovae neutrinos with energies less than 100 MeV) the recoil of the produced electron/positron during their ionization and subsequent emission of atomic electrons (delta rays) is significant especially when the detector makes use of multi-PMT optical modules.

### 2.1.2 Class KM3Detector

In this class we define the detector geometry (the positions and geometrical characteristics of the OMs), as well as the optical properties of the various materials. In the simulation we assume that the OM consists of a group of photocathodes, in order to take into account the case of a Multi-PMT OM. The photocathodes of an OM are defined as sensitive detectors and all the optical photons hitting these are registered in the class KM3SD. The geometry of the neutrino telescope is not coded in the class, but the GDML geometry description markup language is used [9]. This way different detector geometries can be described by editing a simple xml file placing the detector components in the world volume according to the GDML language format.

The setup of the detector in the GDML file should follow some rules. Examples of KM3Sim geometries can be found in the examples distributed with the HOURS_KM3Sim program.

### 2.1.3 Class KM3Cherenkov

This class replaces the original GEANT4 class responsible for the emission of optical photons due to the Cherenkov effect. Since the Cherenkov process is a continuous process, each charged particle with sufficient energy would produce a very large amount of Cherenkov photons even if these photons would never reach the OMs of the detector. As a result the tracking algorithm would spend most of the CPU time for tracking low energy photons. To deal with this problem a method (its application is selectable during compilation) to speed up simulation has been developed. According to this method, only photons that are capable of reaching the detector optical modules are produced (for details how this is done effectively see Appendix B). This means that a photon is tracked only if its direction is going to hit an optical module. However some of these photons are going to be scattered and miss the OM. These photons are discarded in the class KM3OpMie. Moreover to take into account not only direct (unscattered) photons parametrizations have been created that describe the flux of scattered photons per unit length of the charged particle's step as a function of the distance to the emission point and the angle with respect to the particle's trajectory. The application of this method is implemented in class KM3EMDirectFlux (see paragraph below). This way the simulation is speed up without loosing any detectable photon flux.

### 2.1.4 Class KM3SD

In this class we process the hits of the simulation. Every produced optical photon crossing the photocathodes of an OM is registered, taking into account the quantum efficiency and angular acceptance of the photomultiplier. In the following we define such a crossing a hit. For each hit we gather the following information:
1. The id of the cathode
2. The time stamp of the hit
3. The id of the primary particle creating the hit
4. The physical process of the primary particle that resulted in the hit creation
5. The multiplicity of the hit (optional, explained below).
   This information is kept using the container class KM3Hit. For each hit an instance of this class is created.

For event simulations resulting in a large number of hits per cathode there is the option to group the hits in time, in bins of 0.5 ns. In that case the time stamp of the hit is the mean of the arrival times of the photons in the group, while the multiplicity of the hit is the number of photons in that group. In case of such a merging the values of items 3 and 4 are those of the first photon.

### 2.1.5 Class KM3PrimaryGeneratorAction

In this class the type, momentum and initial position vectors of the primary particles are defined. Depending on the functionality mode, these particles are either described in the input EVT file (e.g. for neutrino interaction events or atmospheric muon bundles), or their particle code is input on the command line (e.g. for creating parametrization tables for certain type of particles), or they are of fixed type (e.g. electron or gamma secondaries from $^{40}$K decays or positrons from low energy SN neutrino interaction).

**2.1.6 Class KM3EMShowerModel**

Class for the fast simulation of the response of the detector to electromagnetic showers. The particles creating the showers are electrons/positrons and gammas. The photoelectron output is treated the same for these particles, except for a small position offset of the gamma, along its track. This is to account the distance a gamma travels before its first interaction in the medium. The pion zero electromagnetic showers are not simulated by this class, due to their short lifetime and their decay to two gammas. The parametrization is applied to the world volume (see class [KM3Detector](KM3Detector) )

This class uses tabulated values of the photon output, created during parametrization runs. The tables contain the photon output in bins of lepton energy, distance of the shower to the optical module and orientation of the shower relative to the OM position and direction.

The validity of this parametrization has been checked with a test detector geometry containing several optical modules and is shown in the following plots. The plots show the comparison of the number of photoelectrons produced by using either the fast simulation or the full simulation for an 100GeV electron. The upper left plot shows the photomultiplier orientation relative to the electron. The other comparison plots are versus the shown angle or distance. The upper right plot shows the comparison of the photon arrival time profiles for angle 90 degrees and arbitrary distances of the PMT to the starting point of the shower (the creation point of the electron). The middle plot shows the ratio of the total number of photoelectrons detected versus the angle for fast and full simulation and arbitrary distances of the PMT to the electron creation point. The lower plot shows the ratio of the total number of photoelectrons detected versus the distance of the PMT to the electron creation point for fast and full simulation and arbitrary angles.

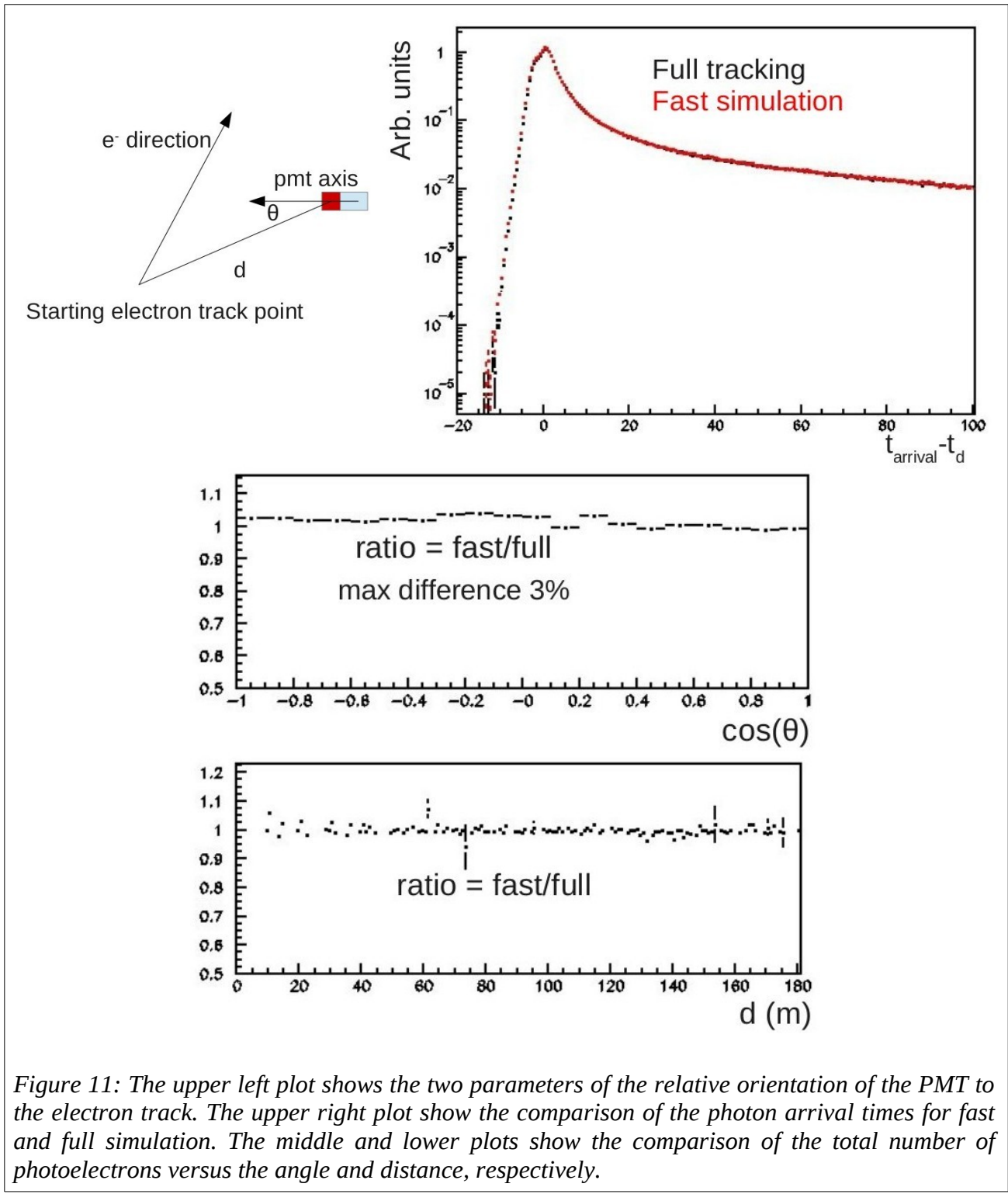

*Figure 11: The upper left plot shows the two parameters of the relative orientation of the PMT to the electron track. The upper right plot show the comparison of the photon arrival times for fast and full simulation. The middle and lower plots show the comparison of the total number of photoelectrons versus the angle and distance, respectively.*

### 2.1.7 Class KM3EMDirectFlux

This is a class storing and interpolating the parameterization tables for the fast simulation of the response of the detector to Cherenkov photons from fast charged tracks. The particle speed should be more than 0.99c. This class is used in KM3Cherenkov when such a particle is encountered. For particles that are not muons, and if the mean number of Cherenkov photons in this particle's tracking step is greater than 20, then the class is used for all photons emitted in this step, scattered and unscattered. If the number of photons emitted from non-muons is small then these photons are explicitly tracked. For fast muons, the parametrization implemented in this class is used only for the scattered optical photons. The parametrization tables have been created only for the scattered photons. The unscattered ones are taken into account by explicitly tracking photons pointing to an optical module while rejecting them if they suffer optical photon scattering. The application of the parametrization for muons is not done step-wise, i.e. used for every tracking step of a muon, but the photons that the muon could emit are grouped together every 50cm of the muon track and for each group the parametrization is applied. The validity of this parametrization has been checked with a test detector geometry containing several optical modules and is shown in the following plots. The plots show the comparison of the number of photoelectrons produced by using either the fast simulation or the full simulation for a muon. The upper left plot shows the photomultiplier orientation relative to the muon track. The other comparison plots are versus the shown angle or distance. The upper right plot shows the comparison of the photon arrival time profiles for angle 90 degrees and arbitrary distances of the PMT to the point of muon track where the most detected photons have been emitted. The middle plot shows the ratio of the total number of photoelectrons detected versus the angle for fast and full simulation and arbitrary distances of the PMT to the muon track. The lower plot shows the the ratio of the total number of photoelectrons detected versus the distance of the PMT to the muon track for fast and full simulation and arbitrary angles.

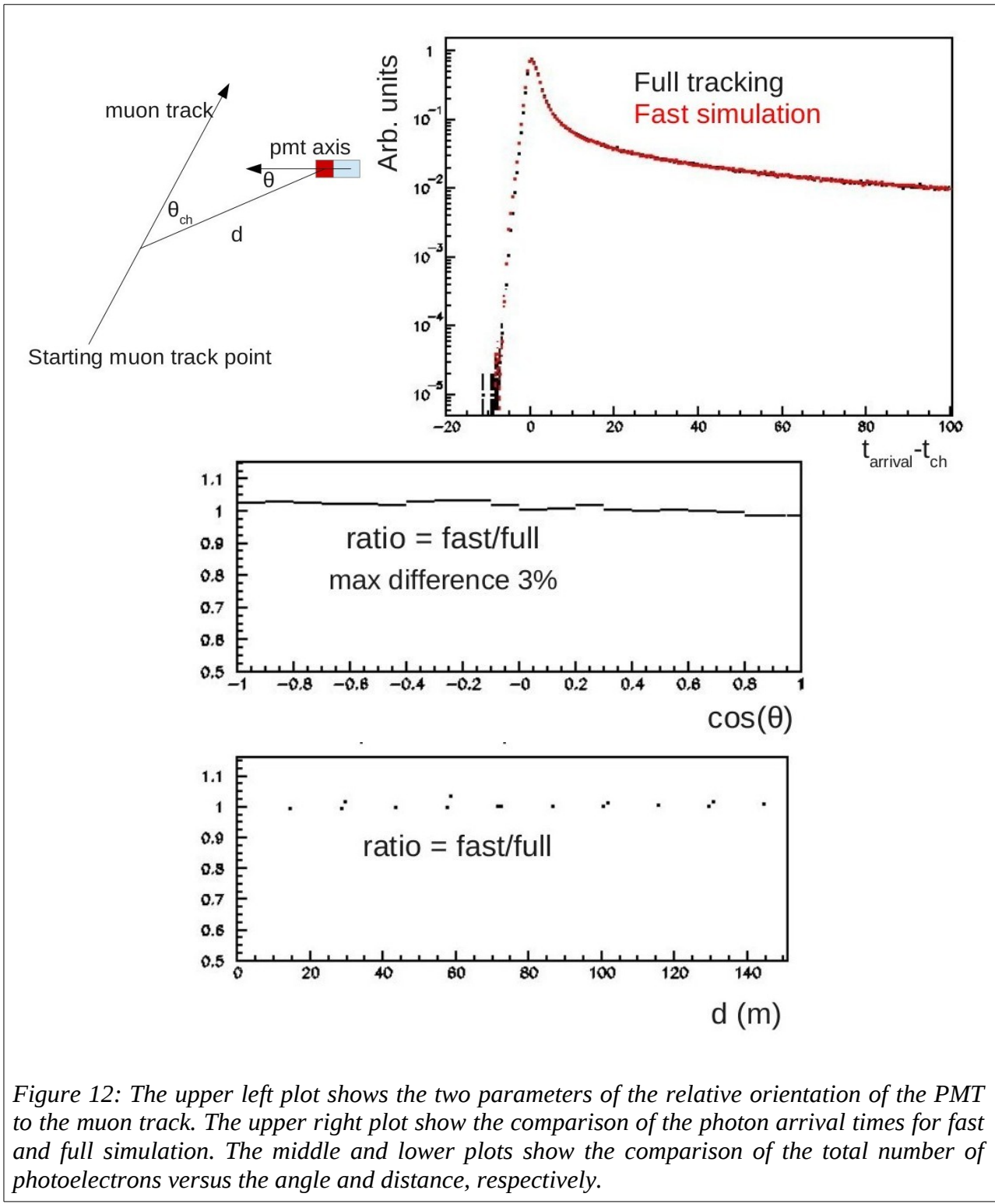

*Figure 12: The upper left plot shows the two parameters of the relative orientation of the PMT to the muon track. The upper right plot show the comparison of the photon arrival times for fast and full simulation. The middle and lower plots show the comparison of the total number of photoelectrons versus the angle and distance, respectively.*

### 2.1.8 Class KM3HAShowerModel

Class for the fast simulation of the response of the detector to hadronic showers. The particles creating the showers are the most common hadrons found in neutrino interactions. These particles are charged pions, charged kaons, kaon zero long, nucleons and antinucleons. Neutral pion and kaon zero short have short life time and decay before they interact in the medium.

Like in the class for the fast simulation of the response of the detector to electromagnetic showers (KM3EMShowerModel), this class uses tables with photon outputs of the hadronic showers. However, hadronic showers produce secondary muons from the decays of pions and kaons. These muons, although they have low energy, due to their long range they produce large longitudinal and lateral fluctuations of the photon output of the hadronic showers. In order to avoid these fluctuations in the table values, the emission of secondary muons from the hadronic showers is suppressed (see section 2.1.10), during the table building. The photon output of these muons are later taken into account during simulation runs, by inclusion of muons as primary particles with the use of the container class HAVertexMuons.

In the fast simulation method applied with this class, we use the approximation that the longitudinal and lateral profiles of the photon flux from the showers emanating from simulated charged hadrons are the same. The absolute flux for each charged hadron has been estimated separately, but the longitudinal and lateral profile used is considered the same. The same kind of approximation is used for neutral hadrons. In the following plots the validity of these approximations are demonstrated. The plots show the comparison of the number of photoelectrons produced by using either the fast simulation or the full simulation for charged pions. The upper left plot shows the photomultiplier orientation relative to the pion track. The upper right plot shows the comparison of the photon arrival time profiles for angle 90 degrees and arbitrary distances of the PMT to the track's starting point. The middle plot shows the ratio of the total number of photoelectrons detected versus the angle for fast and full simulation and arbitrary distances of the PMT to the track's starting point. The lower plot shows the the ratio of the total number of photoelectrons detected versus the distance of the PMT to the track's starting point for fast and full simulation and arbitrary angles.

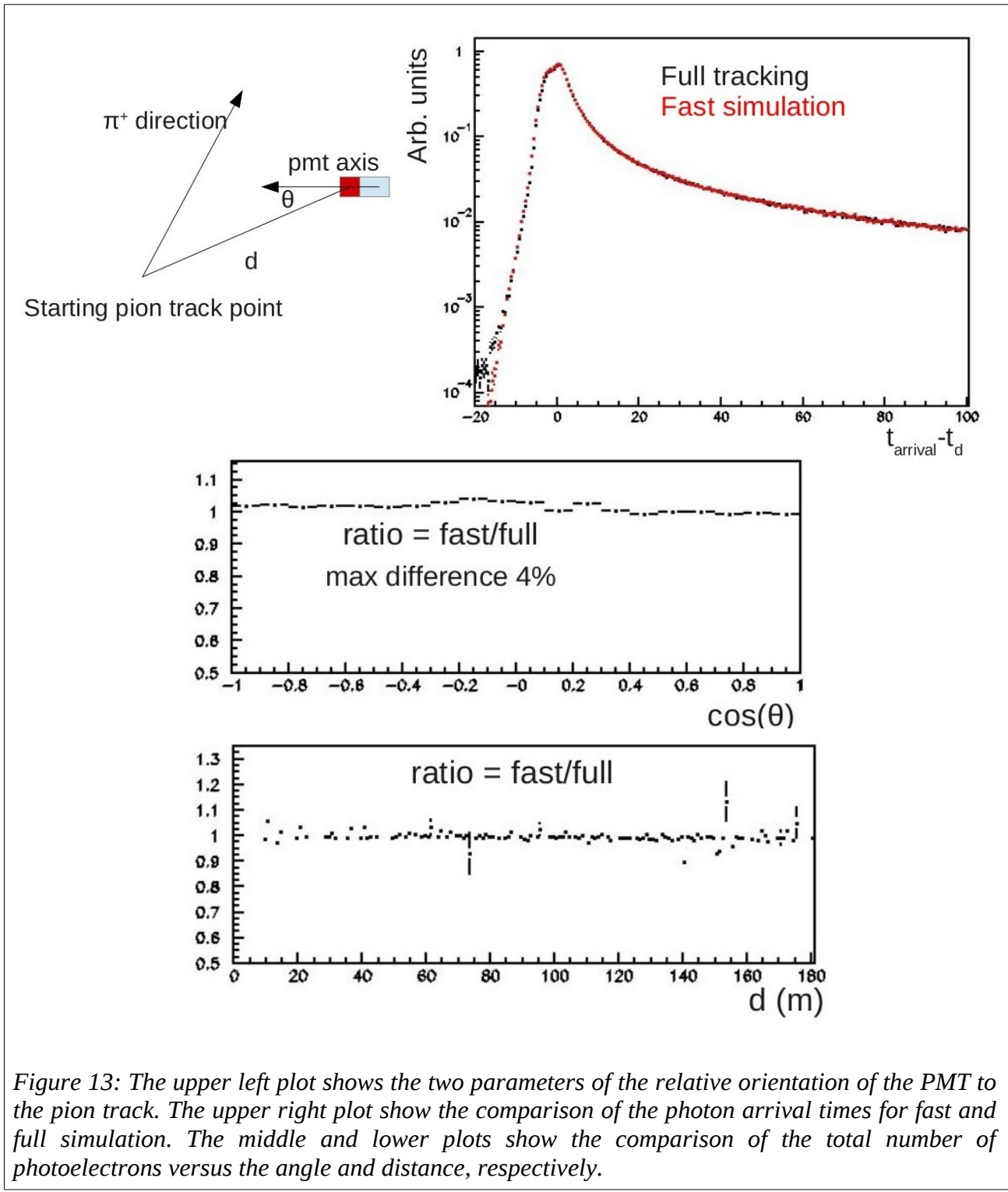

*Figure 13: The upper left plot shows the two parameters of the relative orientation of the PMT to the pion track. The upper right plot show the comparison of the photon arrival times for fast and full simulation. The middle and lower plots show the comparison of the total number of photoelectrons versus the angle and distance, respectively.*

### 2.1.9 KM3EMDeltaFlux

Class storing the parametrization photon flux from low energy electrons (delta rays) and interpolating for the distance. The photon flux stored in the parametrization tables have been normalized, during the table creation, with the total kinetic energy of the produced delta rays, the maximum quantum efficiency of the photocathode and the total photocathode area of the optical module. The distance is the one between the creation location of the electrons and the Optical Module. This class has similar functionality with the classes KM3EMDistanceFlux and KM3EMDirectFlux used for showers or Cherenkov photons from muons, respectively. It is used in class KM3StackingAction for the fast simulation of the detector response to optical photons from low energy electrons created by primary muons due to ionization.

### 2.1.10 KM3StackingAction

Class with actions to be taken when a new particle is created by a physics process during the simulation. These actions are also taken for the primary particles generated in KM3PrimaryGeneratorAction. This a standard GEANT4 user implemented class. Here the actions taken are the following: Electrons below the Cherenkov threshold are disregarded. Gammas below the electron Cherenkov threshold are disregarded. Produced neutrinos are killed. If the particle is not a muon and is created far from the detector (more than a few maximum optical absorption lengths away from the nearest optical module) is disregarded. If the particle is a muon and is not going to pass near the detector is disregarded. When fast simulation of delta rays is used, then the low energy electrons from muon ionization are grouped together and at end of the event are used to produce hits according to the delta ray fast simulation technique implemented in class KM3EMDeltaFlux . For runs producing tables with information of secondary muons produced by the hadronic component of a neutrino interaction, all primary particles that are not hadrons are disregarded. Also the secondary muons from the hadronic component are registered (see class HAVertexMuons ) . For runs with the fast simulation of hadronic showers enabled, all secondary energetic muons are disregarded, since they are included as primaries in KM3PrimaryGeneratorAction using the container class HAVertexMuons .

### 2.1.11 KM3MuonParam

This class is for the acceleration of run time for atmospheric muon simulation. Each muon recorded by the EAS simulator at the sea surface is transported to the surface of the CAN, taking into account all energy losses. The energy losses are not calculated with simulation for this muon, but they are tabulated in PDFs that describe the probability for a muon to have a specific energy, after crossing a certain distance in water. These tables have been created for various distances and initial energies of the muons. Muon multiple scattering is not accounted for, so we consider that the muon transports in a straight line. If the muon has a nonzero probability to be absorbed before reaching the detector's CAN, this is taken into account in the calculation of the event weight. For muon bundles the event weight is the probability to have at least one muon that is capable of reaching the detector.

### 2.1.12 Other classes used

Other classes used are:
- HOURSevtREAD , HOURSevtWRITE for reading input particle information in ANTARES EVT format and writing simulation data to output in the same format.
- KM3Cathods a container class for keeping information about all the photocathodes of the detector.
- KM3EventAction is a class of one of the user's optional action classes. The methods of this class are invoked at the beginning and the end of each event processing. In HOURS_KM3Sim this class is used for registering the information about muon transportation through the detector. It is also used in parametrization tables building mode for initialization of the arrays keeping the tables.

- KM3EMEnergyFlux is a class storing the parametrization photon flux from electromagnetic showers emanating from e+, e-, γ.
- KM3HAEnergyFlux is a class storing the parametrization photon flux from hadronic showers.
- Classes KM3EMAngularFlux, KM3EMDistanceFlux, KM3EMTimePointDis are container classes used by other classes storing parametrized photon flux from various sources.
- HAVertexMuons is a container class to keep secondary muons from hadronic showers. These muons are used when the hadronic shower parametrization is enabled. Refer to paragraph 2.1.7 (Class KM3HAShowerModel).
- KM3Hit is a class that keeps the properties of each hit, i.e. the hitted photocathode, the time of the hit, the initial particle that created the hit, as well as the physics process that created the hit.
- KM3OpMie is a class that applies the particulate (Mie) scattering and the Rayleigh scattering of optical photons.
- KM3SNGarching is a class that produces neutrino interaction events using the electron antineutrino flux from supernovae explosions according to the Garching model.
- KM3SteppingAction is a class inherited from the standard GEANT4 class G4UserSteppingAction. This class is used non-trivially only for primary muons, and through this class muon information is recorded. This information contains the energy and direction of the muon when it enters the detector, as well as decay information when the muon decay inside the detector. Moreover, another functionality of this class is the rejection of muons that can no longer give any signal to the detector.
- KM3TrackInformation is a class used to keep generation information for each particle. This way for each produced photoelectron the physics process and the charged particle that created this photoelectron can be recorded.
- KM3TrackingAction is a class used to pass track information from a parent to a daughter particle, and record information of the secondaries created by muon decay inside the detector instrumented volume.

## 2.2 Running modes and examples

There are several running modes of operation of the HOURS_KM3Sim simulation program. There are four types of modes. The first one is the simulation of high energy events, neutrino interaction events or atmospheric shower events. The second one is the simulation of low energy events, regarding signal or background. The third one is calibration runs, regarding laser or LED runs. And the last one is a category of operation modes for building parametrization tables used for fast simulation or muon energy estimation. Examples of runs with all running modes are included in the distribution. See these examples and the doxygen documentation for more details. In the following all the running modes are briefly described, along with some examples.

### 2.2.1 Neutrino interaction events or atmospheric muon & neutrino bundle events

This running mode can be applied either for the simulation of the detector response to the secondaries of high energy neutrino interaction events in the vicinity of the detector, or to the muons and neutrinos of extensive atmospheric showers. In the first case the events are simulated internally by the generation part of HOURS, using PYTHIA for the creation of secondaries. In the later case the events are initially simulated by CORSIKA, while the generation part of HOURS is responsible for checking if an extensive air shower is capable to give muons or neutrinos with a detectable signal, and for formatting the event in the format used in HOURS software chain.

Depending on the preprocessor commands during the compilation, the simulation according to these running modes can make use of parametrizations of Cherenkov photon flux from electromagnetic and hadronic physical processes in the detector's instrumented volume.

For this running mode there are four examples:

(1) Example "example_NoFastSim", is a run simulating events of neutrino interactions. The input file has been generated by the generator package HOURS_Gen. The fast simulation is not applied in this example, so all particle tracks and optical photons are explicitly tracked by GEANT4. Details:
- The detector consists of 115 vertical arrays (strings) of Optical Modules (OMs). Each array contains 18 OMs with a separation of 36 m between them. The detector's footprint is roughly circular, with a radius of 500m. The mean distance of a string to the nearest one is around 90m. The Optical Module has identical modeling as in the example $^{40}$K run (case 5 below).
- The input file contains 1712 events and it is in ANTARES EVT format. These events are muon neutrino interactions, with neutrino energies ranging from 100GeV to 100PeV. The neutrino spectrum is a power law spectrum with spectral index -1.5. The incident neutrino directions are over all sky, i.e. 4pi solid angle. The number of events tried in the generation step are 10000, but only 1712 of them potentially give signal.

(2) Example "example_FastSim", is a run simulating events of neutrino interactions. The input and the detector geometry is the same as in the example (1), but here fast simulation is applied. The simulation is 100 times faster compared to the simulation without using parametrizations.

(3) Example "example_FastSim_CORSIKA_woNeutrino", is a run simulating the detector's response to atmospheric muon bundles produced by CORSIKA and handled by the HOURS_Gen generator . The produced atmospheric neutrinos from the same shower producing the muons are disregarded. Details:
- The detector geometry is the same as in the example (1), and fast simulation is applied.
- The input file describes the atmospheric muon bundles produced by EAS initiated by iron nuclei in the energy range 30TeV to 1PeV and zenith angles 0 to 87 degrees. The primary spectrum has spectral index -1.5

This running mode has been used for the estimation of atmospheric muon bundle multiplicity and event rate presented in Section 1.3.1.

(4) Example "example_FastSim_CORSIKA_WithNeutrino", is a run simulating the detector's response to atmospheric muon and neutrino bundles produced by CORSIKA and handled by the HOURS_Gen generator . From the bundle of the produced atmospheric neutrinos one is randomly chosen to interact in the detector's instrumented volume. The neutrino interaction event weight has been calculated taking into account path length, matter density and interaction cross sections. The neutrino interaction and its secondaries are simulated with the HOURS_Gen generator. Details:
- The detector geometry is the same as in the example (1), and fast simulation is applied.
- The input file describes the atmospheric muon and neutrino bundles produced by EAS initiated by proton in the energy range 1TeV to 10EeV and zenith angles 0 to 87 degrees. The primary spectrum has spectral index -1.5. In the input file only events with atmospheric muons able to reach the detector are included.

This running mode has been used for the evaluation of the performance of a technique to reduce the background of atmospheric neutrinos when observing the diffuse flux of high energy cosmic neutrinos as it is explained in Section 1.3.3.

### 2.2.2 Simulation of low energy events

In this running mode the primary particles are either an electron or gamma from the $^{40}$K decays or a positron from the interaction of an electron antineutrino from a supernova with a proton of the water medium. The $^{40}$K decays to Calcium 40 (branching ratio 89.338%) with the emission of an electron (with maximum energy 1.33MeV) and an antineutrino, or it decays to Argon 40 with the emission of a gamma with energy 1.460MeV.

#### 2.2.2.1 Optical background simulation of $^{40}$K decays

Potassium 40 is the major source of optical background in underwater neutrino detectors. Even if the signal rate from $^{40}$K radioactive decays are many order of magnitudes greater than that from energetic particles crossing the detector's volume, with the use of coincidence triggers this noise can be significantly reduced. Moreover, the $^{40}$K decays provide a continuous and very well known optical photon source for online monitoring of the optical module performance. The simulation of the response of the detector's optical modules to $^{40}$K decays is significant to evaluating the performance of the triggering techniques. In the running mode of HOURS_KM3Sim for $^{40}$K simulation the detector consists of a single optical module, while the $^{40}$K decays within a spherical volume around this optical module.

For this running mode there is one example:
(5) Example "example_K40Runs", is a run for $^{40}$K decays. Details:
- The detector geometry used is a single Optical Module at the center of the coordinate system. The Optical Module consists of 31 3-inch photomultipliers. The photomultipliers are modeled in the GDML file (see section 2.1.2) as thin disks with radius 4.7462cm. The enlarged radius of the PMT is to account for the light concentrating ring build around the photocathode.
- The decay's positions are random within a 2m radius sphere centered at the center of the OM.

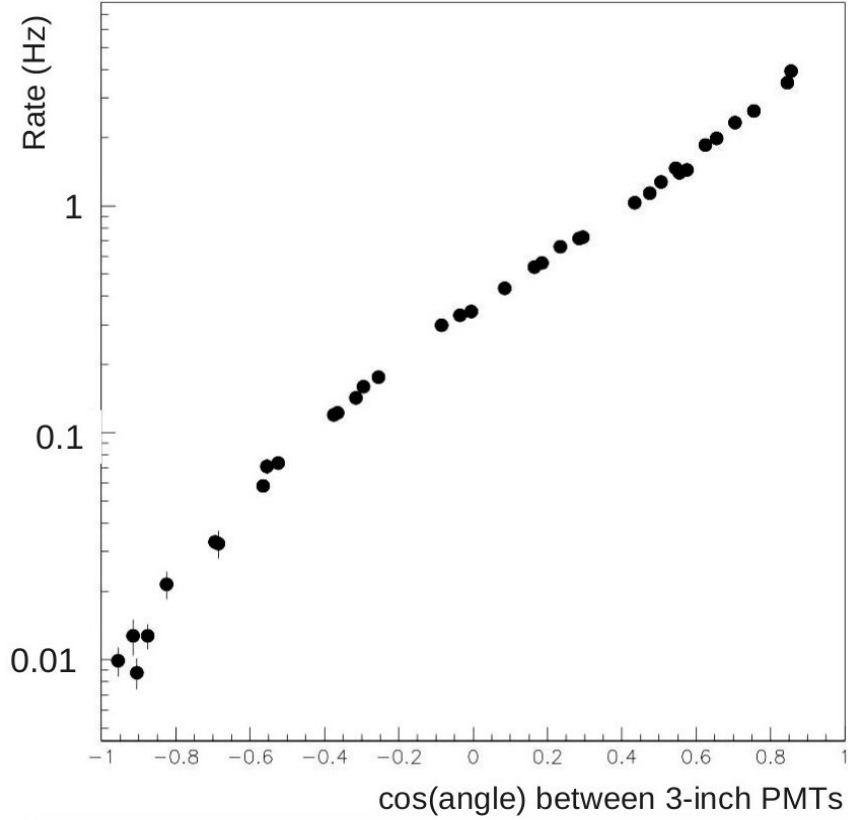

*Figure 14: The genuine double coincidence hit rate on two small PMTs as a function of the orientation angle between them. The rate has been estimated by producing billions of $^{40}$K decays around an Multi-PMT optical module, with the decays' positions within a 10m radius sphere centered at the optical module. It has also been estimated that the double coincidences caused by decays at a distance more that 10m from the optical module is less than 1% of the overall coincidence rate.*

An important aspect of the $^{40}$K decays in optical modules with many small photomultipliers is the rate of genuine double or more coincidences between two or more small pmts. These coincidences are hits on the optical module that contain two or more activated pmts in a small time window of 10ns. These hits are created by the same $^{40}$K decay. By knowing these rates the performance of various triggering and noise reduction algorithms can be estimated. Fig. 14 shows the genuine double coincidences of an optical module consisting of 31 3-inch photomultipliers. This optical module is the one developed and adopted for KM3NeT.

**2.2.2.2 Simulation of low energy neutrino interactions from Supernovae.**

The majority of the energy carried out during the first phases of a supernova explosion comes in the form of neutrinos. These neutrinos with energies a few tens of MeV can interact with the water medium of the underwater neutrino telescope and give a possibly detectable signal. The most relevant detection channel for these neutrino energies is the interaction of electron antineutrinos with the proton nuclei of the water medium via the quasielestic reaction (inverse beta decay):

$$\bar{\nu}_e\, p \rightarrow e^+ n$$

The emitted positron has enough energy to emit Cherenkov photons and thus give a signal. These positrons produce many mini electromagnetic showers that span to about a few centimeters each. These EM showers cannot be separately identified due to their small size, but since the neutrino flux is enormous they can give a rise to the photomultiplier counting rate. The simulation of such a source is significant for the evaluation of the performance of an underwater neutrino telescope to detect supernovae explosions. In the running mode of HOURS_KM3Sim for SN neutrino simulation simulation the detector consists of a single optical module, while the neutrinos interact within a spherical volume around this optical module.

For this running mode there is one example:

(6) Example "example_SNRuns", is a run for SN neutrino detection. Details:
- The detector geometry used is a single Optical Module as in the example (5) above
- The neutrino flux is according to the Garching model . References:
  - L. Hudepohl et al, Phys. Rev. Lett. 104,251101 (2010) (also in https://arxiv.org/abs/0912.0260 ; http://dx.doi.org/10.1103/PhysRevLett.105.249901 ; http://dx.doi.org/10.1103/PhysRevLett.104.251101 )
  - M.T. Keil, et al, The Astrophysical Journal, 590:971-991, 2003 (also in http://arxiv.org/abs/astro-ph/0208035 ; http://dx.doi.org/10.1086/375130 ).
- The zenith and azimuth angles of the SN source is random for each event. The position of the SN in the sky is not expected to affect the signal much, since the optical module described has an almost uniform angle acceptance and the mini EM showers produce a more or less isotropic flux of Cherenkov photons.
- The distance of the SN is 10 kpc, and the time integration upper limit for the fluence is 2.95 sec .
- The interaction positions are random within a 5m radius sphere centered at the center of the OM

Any detection technique of a supernovae using the neutrino channel must account for the optical noise from $^{40}$K and the noise from atmospheric muons. In the sea water the optical noise from $^{40}$K decays is many order of magnitudes greater than the signal from a nearby supernovae. However, the use of Multi-PMT optical modules consisting of many small PMTs can be of great assistance by exploiting the genuine coincidences between activated PMTs on the same OM. The positrons emitted by the SN neutrino interactions have energy at least one order of magnitude greater than the electrons emitted by the $^{40}$K decays. This means that supernovae neutrino interactions will give signal with more activated PMTs of the same OM. The multiplicity of the coincidences from SN neutrinos will be higher compared to that of $^{40}$K noise. However, atmospheric muons also create hits with many activated PMTs on the same OM. Fig. 15 presents the expected number of activated OMs in an observation time interval of 2.95sec for SN neutrino interactions, $^{40}$K decays and atmospheric muons. Muons also activate nearby OMs, while each SN neutrino interaction or $^{40}$K decay can give signal only to a single OM. Using an appropriate analysis that exploits these facts it has been shown that an underwater neutrino telescope with 6160 Multi-PMT OMs can unambiguously detect (>5σ) a supernova at a distance up to 23 kpc from Earth.

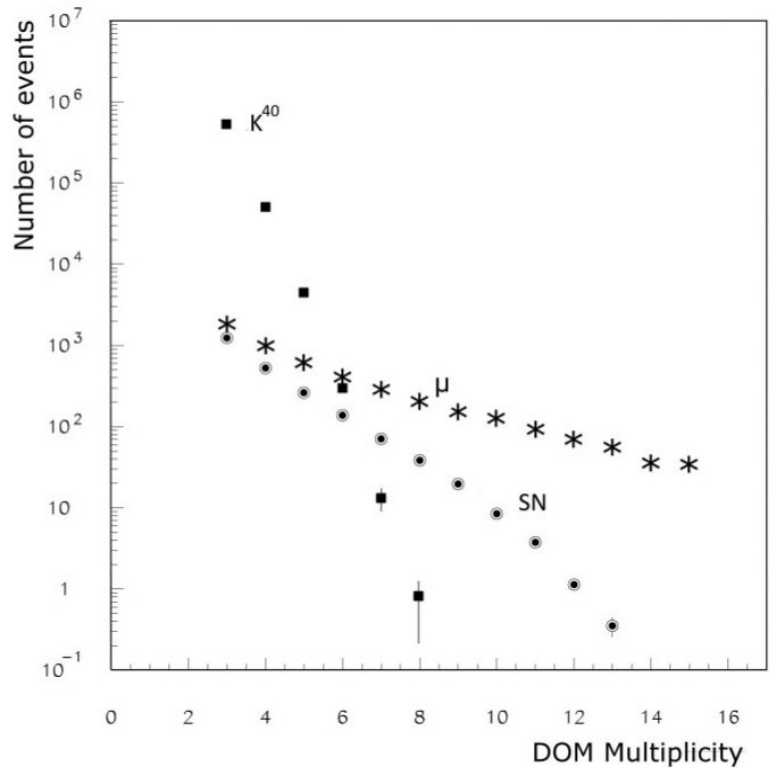

*Figure 15: The expected number of observed events (active OMs), in an observation time interval 2.95 s, as a function of the OM multiplicity for three type of events: $^{40}K$ decays, atmospheric muons and SN neutrinos. The SN explosion is assumed to occur at a distance of 10 kpc and the neutrino telescope is assumed to contain 6160 OMs.*

### 2.2.3 Calibration Runs

Calibration run mode in the simulation exist for Laser beams and LEDs. For this running mode there is one example:

(7) Example "example_LaserRuns", is a run simulating a single optical module response to a laser beam. Details:
- The detector geometry used is the one that is also used for the example $^{40}K$ decays run (example 5). A single optical module at the center of the coordinate system.
- The run is for a Laser at position (0,1,-10)m pointing parallel to the unit vector (0,0,1).
- The Laser's wavelegth is 400 nm.

Using such a calibration run and a single multi-PMT optical module (like the one developed for KM3NeT) the performance of a technique for the estimation of the optical properties of the water medium can be studied. The technique estimates the optical scattering parameters of the water medium by comparing the number of detected photons from different PMTs of the same optical module. The photons are emitted by a Laser and although the Laser is not pointing to the PMTs, some of these photons are detected due to scattering. The number of detected photons depends on the optical scattering parameters. An example is presented on Fig. 16. The plots show the distributions of the arrival times of the detected photons by a PMT when several Laser pulses are emitted. The scattering model used is a combination of Mie (particulate) scattering and a modified Rayleigh scattering :

$$\frac{dP}{d\Omega_s} = F(cos\theta_s; p, a_{Rayl}, a_{Mie}) = p \times g(a_{Rayl}, cos\theta_s) + (1-p) \times f(a_{Mie}, cos\theta_s) \quad,$$

where F is the phase function, $\theta_s$ is the scattering angle, p is the relative contribution of the Rayleigh scattering, g is the Rayleigh scattering phase function and f is the Mie scattering phase function:

$$g(a_{Rayl}, cos\theta_s) = \frac{(1 + a_{Rayl} cos^2\theta_s)}{4\pi(1 + \frac{1}{3}a_{Rayl})}$$

$$f(a_{Mie}, cos\theta_s) = \frac{1}{4\pi} \frac{(1 - a_{Mie}^2)}{(1 + a_{Mie}^2 - 2 a_{Mie} cos\theta_s)^{\frac{3}{2}}} \quad.$$

The parameters $\alpha_{Mie}$ and $\alpha_{Rayl}$ determine the shape of the distributions, and in the case of Mie scattering $\langle cos\theta_s \rangle = a_{Mie}$, while in the case of Rayleigh scattering $\langle cos\theta_s \rangle = 0$. The scattering length do not affect the phase function.

As shown in Fig. 16, the arrival timing distributions depends strongly on scattering length, relative contribution of Rayleigh scattering, p, and mean cosine of scattering angle for Mie phase function, $\alpha_{Mie}$, while do not depend so much on the shape parameter of Rayleigh scattering phase function, $\alpha_{Rayl}$. The dependence of the distributions on the scattering length is similar for all the PMTs, since the scattering length do not affect the distribution of the scattering angles (the phase function). However the dependence on the other parameters is a function of the PMT position and orientation relative to the Laser beam.

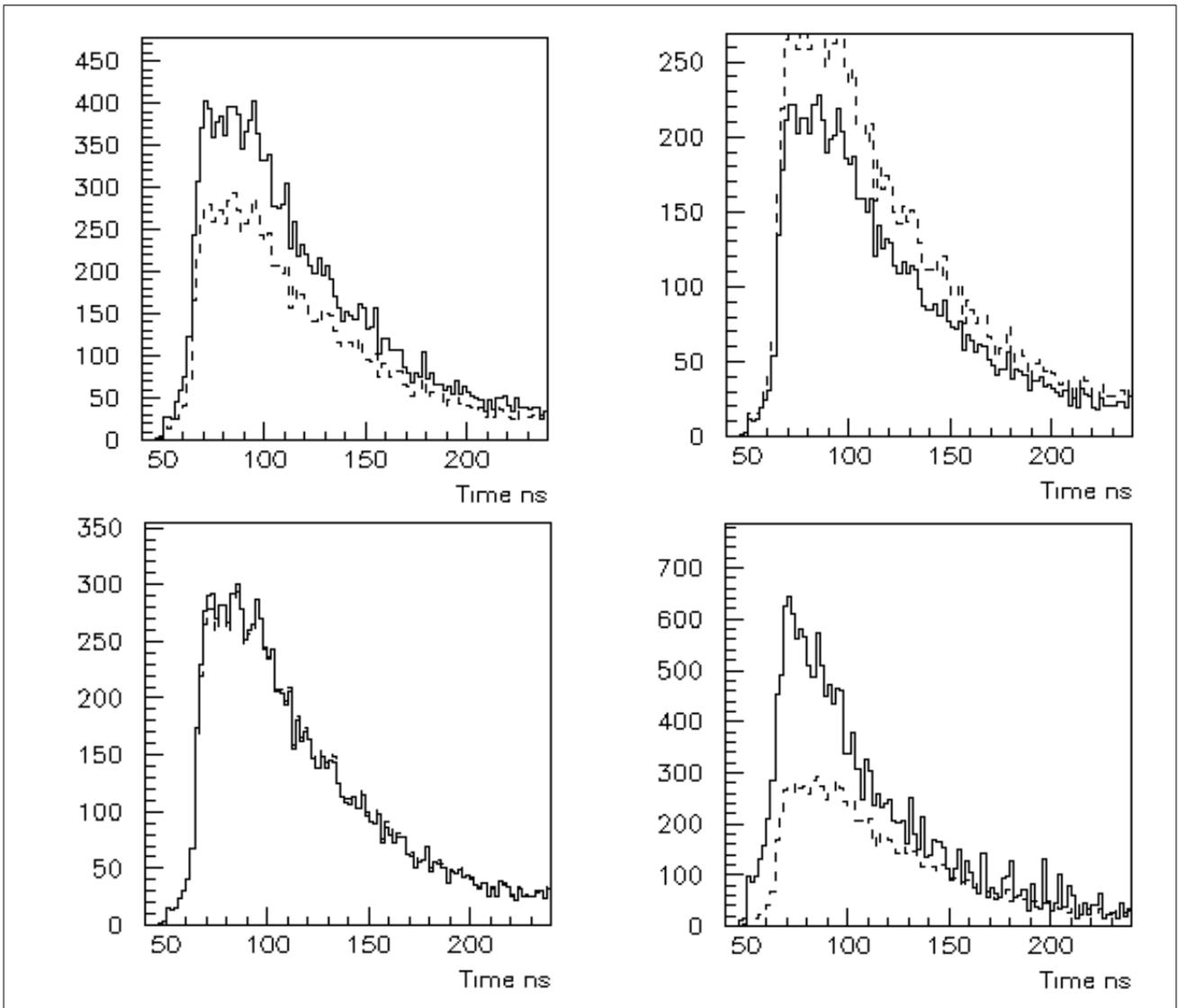

*Figure 16: The dependence of the arrival time distributions of detected photons on several parameters of the optical photon scattering. The upper left plot shows the difference in the distributions for two different scattering lengths, 18m (solid line) and 20m (dashed line). Decreasing scattering length results in more scattered photons. The upper right plot is for two different Rayleigh contributions, p=0.153 (solid line) and p=0.17 (dashed line). Increasing the Rayleigh contribution factor more photons are scattered in larger angles and detected by PMTs away from the Laser's beam path. The lower left plot is for two different Rayleigh contribution shape factors, $a_{Rayl}$=0.77 and 0.853. The indifference of the distributions is due to the PMT's orientation. The lower right plot is for two different means of the cosine of scattering angle for Mie phase function, $\alpha_{Mie}$=0.83 (solid line) and 0.924 (dashed line). Increasing the $\alpha_{Mie}$ results in smaller scattering angles and less photons are detected from PMTs that are not oriented along the beam's path.*

### 2.2.4 Parametrization table building modes

There are four type of running modes for the creation of parametrization tables. These tables are later used in other running modes when fast simulation is enabled, or are used for muon energy fit. The modes are the following.

#### 2.2.4.1 Runs for building parametrization tables of electromagnetic processes

This running mode is used in order to build tables for fast simulating most common electromagnetic processes. These processes are the ones that produce many Cherenkov photons, and by parametrizing the photon output of these processes speeds the simulation a lot. Comparison of simulation runs with or without using these parametrization tables are described in sections 2.1.6 and 2.1.7. The processes whose the photon output is tabulated are:

- Electromagnetic showers initiated by electrons, positrons and gammas. The pion zero electromagnetic showers are not simulated by this class, due to their short lifetime and their decay to two gammas. The longitudinal and lateral profile of EM showers initiated by e-/e+ or gamma is considered the same, except for a shift along the shower axis. This shift depends on shower energy (particle's energy) and is the distance of the shower maximum to the particle's creation point. The photon output of the shower is parametrized in a coordinate system where the zero is not the particle's creation point, but the position of the shower maximum. This position is the same for electrons and positrons but it differs for gammas, due to the distance covered by the gamma before its first interaction. There is one example of table creation for electromagnetic showers:
  - (8) Example "example_BuildEMParamTables", is a run for table creation fully simulating many electron showers. This example run by itself does not have enough statistics to produce fully the parametrization tables. Details:
    - The detector consists of 40 spherical concentric shells with 4 cm thickness. The shells' radii are in the range from 0.5m to 1000m . The electron starts at the center of the shells with an offset depending on the electron energy (see above).
    - The mean photon output of 10 simulated showers each emanated from a 1.0 GeV electron is output.
- Cherenkov photon emission from low energy electrons from muon ionization. When fast simulation is enabled the electrons from ionization of primary muons are not tracked by GEANT4 if their energy is less than 31.6 MeV. The contribution of these electrons to the photon output of the event is taken into account using parametrization tables describing this photon output. These electrons are grouped together in 1m stepping intervals of their parent muon, and for every step the parametrization is applied.
- Cherenkov photon emission from primary muons. For every centimeter a muon travels in water about 300 Cherenkov photons are emitted in wavelengths detectable by a photomultiplier. The tracking of each photon consumes a lot of CPU resources. The application of a parametrization to describe this photon flux at the position of the optical module can greatly speed up simulation. There is a running mode to create the photon flux tables for such a parameterization. The primary particles for such a running mode are optical photons with an energy spectrum and direction that of Cherenkov photons emitted from charged particles moving near the speed of light (>0.99c).

#### 2.2.4.2 Runs for building parametrization tables of hadronic showers

Hadronic showers produce Cherenkov photons, though not as many as electromagnetic showers. Also in contrary to the electromagnetic showers the hadronic ones present more fluctuations in their photon output mainly due to secondary muons produced through the pion and kaon decays. In order to avoid these fluctuations in the parameterized photon output of hadronic showers, the secondary muons from these showers are treated separately. When building the parametrization photon flux tables for hadronic showers all secondary muons with energy above 1GeV are disregarded. When the fast simulation is enabled and these tables are used, in order to take into account the photons from such secondary muons, we create them as primary particles. The number of these muons, their energy and their relative position and direction with respect the starting point of the shower are read from a file (using the class HAVertexMuons) containing such information from many hadronic showers with energies from 100GeV up to 100PeV. This file is build using the appropriate running mode. So the parameterization of hadronic showers involves two running modes:
- A running mode for the creation of the photon flux tables of hadronic showers created from single hadrons. The hadrons that are described by this parametrization are the most frequent ones in high energy neutrino interactions. These are pions, charged kaons, kaon zero long, protons, antiprotons, neutrons and antineutrons. As explained in section 2.1.8 the parameterizations are build for two kings of hadrons, charged ones and neutral ones. The first type is considered to be described by pions, while the second type is considered to be described by kaon zero longs. For this running mode there is one example:
    - (9) Example "example_BuildHAParamTables", is a run for hadronic shower photon table creation, and it fully simulates the shower produced by pions plus. This example run by itself does not have enough statistics to produce fully the parametrization tables.
        - The detector geometry used is the same as in example (8) for the creation of photon flux tables of electromagnetic showers.
        - The simulated pions plus has an energy of 1GeV.
        - The pions starts at the center of the shells with an offset in order for the shower maximum to be at the center of the coordinate system (see example (8)).
- A running mode for the creation of tables of secondary muons from hadronic showers. For the creation of the tables neutrino interactions are simulated using the generation program HOURS_Gen. The electromagnetic component of the secondaries of the interactions are disregarded, so the primaries for this running mode is composed solely from hadrons, the most common are the ones treated by the hadronic fast simulation (see the previous running mode). There is one example for this running mode:
    - (10) Example "example_BuildHAMuonParamTables", is a run creating a list of secondary muons from hadronic showers produced by electron neutrino interaction events. No Cherenkov photons are created, since we only want secondary muons. This example run by itself does not have enough statistics to produce a sufficient sample of secondary muons from hadronic interactions. Details:
        - The detector geometry used is the same as in example (1).
        - The number of neutrino interactions are 1000, with the neutrino having energies ranging from 100GeV to 100TeV.
        - The neutrino spectrum is a power law spectrum with index -1.0.

In section 2.1.8 there is a description of the comparison of simulation runs with or without using the parametrization tables for the photon output of a hadronic shower, as well as the tables with the secondary muons from the same shower.

**2.2.4.3 Runs for building parametrization tables for muon energy fit**

The energy estimation of low energy (<1TeV) muons is based on measuring the track length of the muon inside the detector, since the number of Cherenkov photons emitted by the muon is not depending strongly on energy for such energies (for such energies the main energy loss process for muons is ionization). For high energy muons, since their track length (>1km) exceeds that of the detector's dimensions such an estimation is not feasible. However, for muons with energies greater than 1TeV the number of the produced Cherenkov photons are linear with energy due to discrete radiation processes (bremmstrahlung, pair production and muon photonuclear interaction). So the energy estimation for high energy muons can be based on the number of photons detected by the photomultipliers of the detector. The expected number of photons versus the energy of the muon for various distances of an optical module to the muon track is compiled in tables using a running mode of HOURS_KM3Sim. In this running mode the fast simulation is used for muons passing through a detector consisting of many optical modules in various distances from the muon track. Not all photons emitted by the muon are counted in the tables. The counted photons are the ones having arrival times in the interval [-20ns, +100ns] with respect the arrival times of the photons emitted directly from the muon track. Limiting the timing of the expected photons results in reducing the number of photons in the estimation of the muon energy, and thus reducing the statistics and the accuracy of the energy estimation. However this is necessary, since by opening up the interval too much we also count a considerable number of random $^{40}$K noise in the energy estimation, and this also reduces the accuracy of the energy estimation. The chosen interval is the optimal so that not too many useful photons are lost, and not too many noise photons are counted.

There is one example for this running mode:

(11) Example "example_BuildMuonFitTables", is a run showing how to make full simulation of muons in order to build the tables used for estimating muon energy. Details:
- In this example there are 10 events each with a muon minus with energy 10TeV. The muon starts at position (0,0,-600)m and moves towards the positive z-axis. The stating position is hard coded in method KM3PrimaryGeneratorAction::GeneratePrimaries() .
- The detector geometry contains 725 vertical strings of optical modules in various distances from the z axis, where the muon travels. The distances range from 10m and up to 500m. Each string contains 21 optical modules with a vertical distance 50m between them. The optical module contains one photocathode of 13 inch in diameter.
- Fast simulation of electromagnetic and hadronic showers is enabled.

The shape of the distributions of the number of photons emitted by a muon and detected by the photocathode does not resemble a Poisson distribution, due to the fluctuations of the discrete radiation processes of the muons. In fact the shape of the distribution depends not only from water properties but also from the muon's energy, due to the fact that the relative contribution of each radiation process depends on muon's energy. Due to the varied shapes of the distributions it is not possible to fit each distribution globally, so the full shape of the distributions is tabulated. In Fig. 17 the distribution of the number of photons emitted by a muon and detected by an optical module is presented and compared with a Poisson distribution with the same mean value.

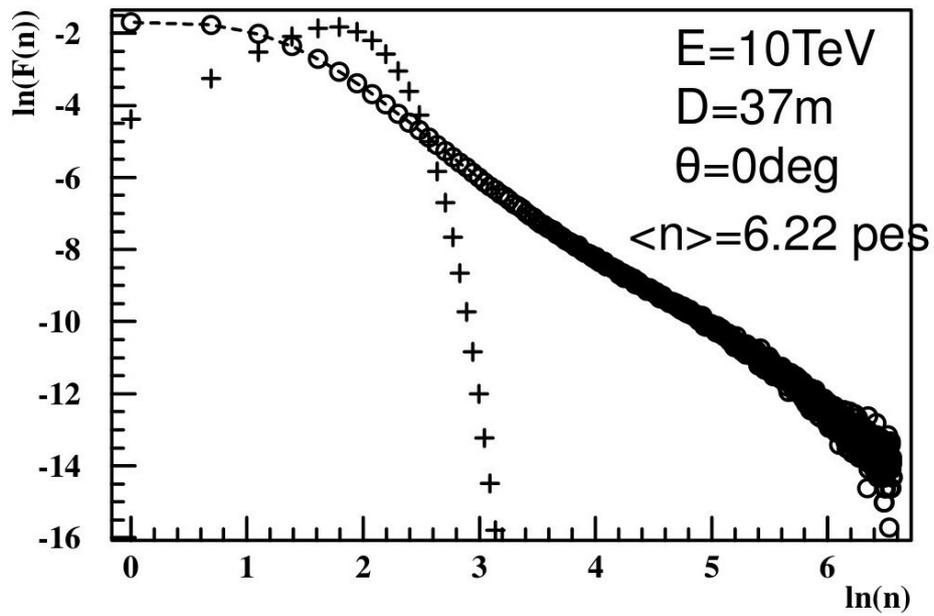

*Figure 17: The distribution of the number of photons, n, emitted by a 10TeV muon passing 37m from a PMT (open circles). The orientation of the PMT (zero degrees in the plot) is relative to the direction along the line connecting the most probable emission point of the muon track to the PMT. Zero degrees means that the PMT is oriented towards the most probable emission point. The crosses present the Poisson distribution with the same mean (6.22) as the actual one.*

# Chapter 3 – Optical Module and optical noise background simulation

This chapter describes the software for the generation of the analog signal from the photoelectrons emitted from the photocathode of an underwater neutrino telescope's photomultipliers. The simulation includes the standard single photoelectron shape (class PulseShape), the charge resolution of the photomultiplier (class PulsePDF), the transit time spread distribution (class PulseTTS), as well as potential preamplifier effects (applied in class HitPMT). The information of the analog signal registered in the output consists of its charge, Time over Threshold (ToT) and the timing of the pulse. The charge of the analogue signal is in single photoelectron units, while the timing of the pulse and its ToT are defined relative to a predefined threshold which is also in single photoelectron units.

The Quantum Efficiency (QE) versus wavelength has been included in a previous step of the simulation, and in this program there is only the possibility to rescale globally the QE to a lower value relative to the one used in photoelectron creation. This can be done also to take into account a possible collection efficiency that has not be included in the photoelectron creation and also a possible shadowing effect from mu-metal cage around the photomultiplier. The program HOURS_KM3Sim that produces the photoelectrons is not saving in its output file information such as incident photon wavelength, direction and point on the photocathode, so any scaling effect to the input photoelectrons in the present program can only be global.

## 3.1 Running modes

The analysis run modes of this program are either creation of hits from signal only (i.e. excluding optical noise background), either from background only (optical background is created in class K40MultiPENoise), or from signal and background. The hit creation from background only is used to estimate the rate of pure background events that pass any predefined trigger algorithm. The program applies simple triggers to the event using the produced hits. The triggering algorithms are based on the number of PMT hit coincidences in an optical module and on the number of OM hit coincidences in neighboring Optical Modules.

## 3.2 Program structure

There are five utility classes in this package. The class PmtContainer describes the detector geometry, the class K40MultiPENoise produces the optical background noise, the class PulseTTS describes the photomultiplier timing resolution on the single photoelectron level, the class PulsePDF describes the photomultiplier charge resolution, while the class PulseShape describes the photomultiplier single photoelectron pulse shape.

For each photoelectron or group of synchronous photoelectrons read from the input file an instance of the class HitPE is created. The main objective of this class is to provide the amplitude of this single/multiple photoelectron pulse at any time instant.

All the HitPE instances for each photomultiplier are accumulated in an instance of the class HitPMT. In this class the possible time overlay of different HitPE instances is taken into account and the PMT hits are produced from the PMT waveform. For each PMT hit the charge, the ToT and the time is registered, while there is also the possibility to deform the photomultiplier pulse due to preamplifier effects.

All the HitPMT instances for each optical module are accumulated in an instance of the class HitOM. In this class all the PMT hits of this optical module's photomultipliers that are in a time interval of 10 ns (adjustable from input) are grouped together and this group defines an Optical Module (OM) hit. An OM hit except from the information inherited from the PMT hits (timing, ToT and charge) it also has the attribute of directionality. The direction of an OM hit is the average direction of the active photomultipliers in the group of the PMT hits. The average is taken using as weight the estimated charge of each PMT hit.

In the following each of the above mentioned classes are explained in more detail.

### 3.2.1 Class PmtContainer

This class reads the detector geometry as it is output from the simulation package HOURS_KM3Sim, it groups the detector's PMTs in groups of Optical Modules and it finds various geometrical characteristics of the detector. The methods of this class are used through out the simulation for accessing specific information about the detector structure. More details about these methods can be found in the doxygen documentation.

### 3.2.2 Class K40MultiPENoise

A class to provide optical noise background to events from $^{40}$K radioactive decays. This class describes the background for the multi-PMT optical module used in the latest design of the KM3NeT. The background consists of:

- Single photoelectron hits on a random PMT of a random optical module. These hits are created from $^{40}$K decays located far away enough from the optical module so that the probability for more than two photons to arrive at more than two PMTs of the same optical module is negligible. The rate of these single hits is read from user input.
- Double photoelectron hits on two PMTs of a random optical module. These hits are created from $^{40}$K decays located close enough from the optical module so that two or more photoelectrons from the same decay can activate two PMTs of the same optical module. The distance from the decay position to the optical module for these kind of hits do not exceed 10 meters. The correlation of the two PMTs is taken into account using a function that describes the double hit probability as a function of the angular deviation between the two PMTs. This probability depends on the angular acceptance of each PMT and it is set for the angular acceptance of the PMTs used in the latest design of the KM3NeT. The rate of the genuine double coincidences versus the relative orientation angle of two KM3NeT PMTs on the same Optical Module can be seen on Fig. 14.
- Multiple (>2) photoelectron hits on three or more PMTs of a random optical module. These hits are created from $^{40}$K decays very close (<2m) to the optical module and consist of three or more photoelectrons created from the same decay activating three or more PMTs of the same optical module. Since there is no practical way to do the same parametrization for multiple hits as it is done for double hits, these multiple hits are described by samples of photoelectrons created from $^{40}$K decays simulated with HOURS_KM3Sim. These samples are kept in file which is read when the instance of the class is created. These samples depend on the angular acceptance of the PMTs and their relative orientation.

Each type of background is created by calling a specific method of the class.

### 3.2.3 Class PulseTTS

A class to provide the application of the PMT Transit Time Spread (TTS) resolution to the photoelectrons. The timing of the photoelectrons that are read from the input are the timing of the incident photons. This class describes the spread of the arrival time of the photoelectrons to the first dynode of the PMT. This spread is considered to be global and not to depend on the point of incidence of the initial photon on the photocathode.

The TTS is modeled as a Gaussian distribution or as an arbitrary distribution. In the first case the user provides the sigma of the Gaussian distribution, while in the later case he provides the full distribution in binned format. The bin values are read from a file. The Fig. 18 shows the PMT TTS of the KM3NeT photomultipliers as measured by the collaboration and used in the simulation program.

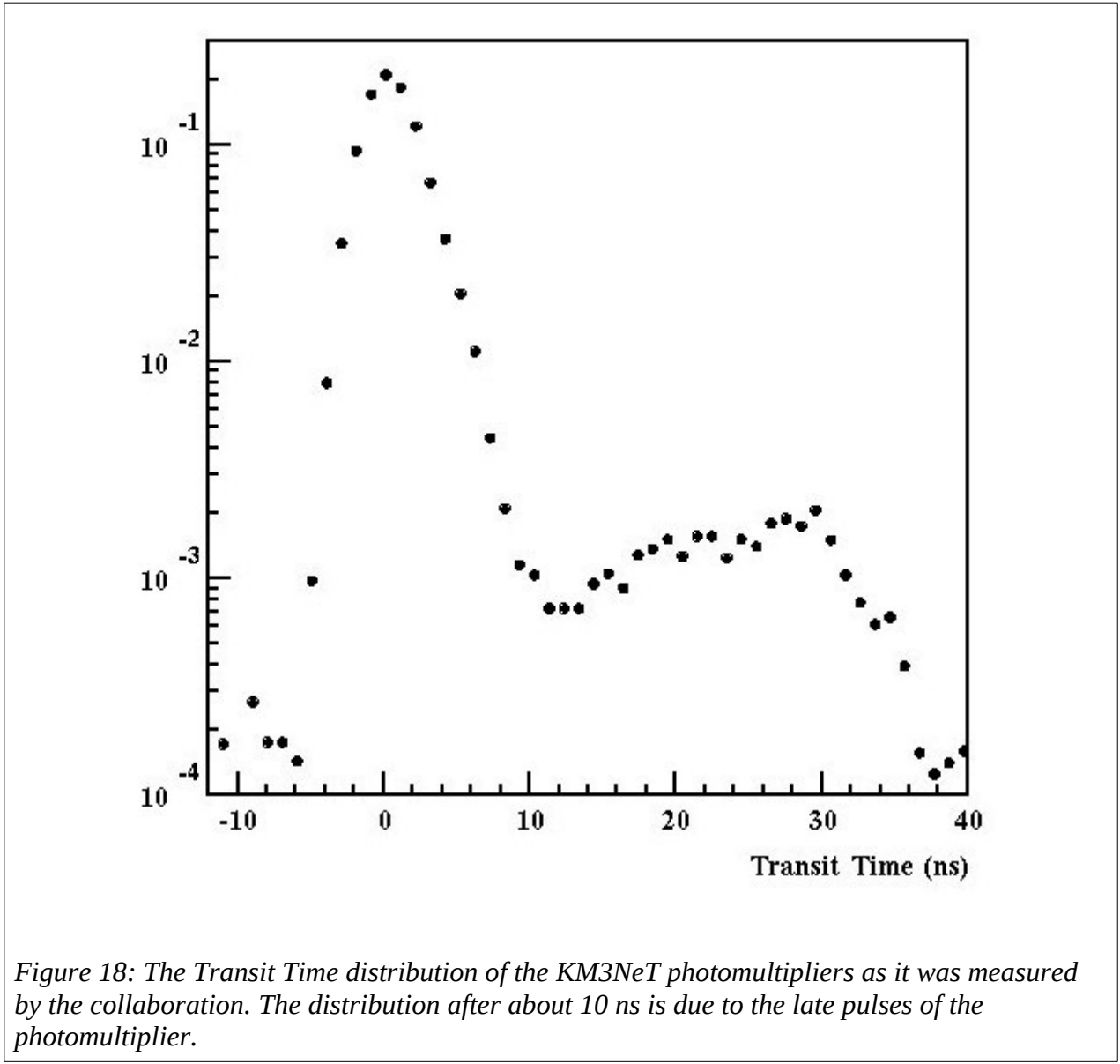

*Figure 18: The Transit Time distribution of the KM3NeT photomultipliers as it was measured by the collaboration. The distribution after about 10 ns is due to the late pulses of the photomultiplier.*

Except from the constructors – one for Gaussian TTS and another for arbitrary TTS – the only other public method of this class is to provide the random time values to be added to the creation times of the photoelectrons.

### 3.2.4 Class PulsePDF

This is a class for the simulation of the charge resolution of the PMT. This is the spread of the charge deposited on the photocathode, when one photoelectron is created from the cathode. In the program this spread is normalized with the mean value of the deposited charge. So the charge resolution used is a percentage of the single photoelectron charge.

The charge resolution in the program is modeled as a Gaussian distribution or as an arbitrary distribution. In the first case the user provides the the sigma of the distribution normalized with its mean value, while in the latter case he provides the full distribution in binned format. The bin values are read from a file. Except from the constructors – one for Gaussian charge resolution distribution and another for arbitrary – the only other public method of this class is to provide the random value that represents the anode charge in single photoelectron units.

### 3.2.5 Class PulseShape

This is a class that provides the pulse shape of the single photoelectron. The anode waveform when a single photoelectron is created on the photocathode is considered to be constant, while the synchronous creation of many photoelectrons creates a waveform that change only in scale. This supposes linearity of the photomultiplier, which is a good approximation for the most common photomultipliers and pulses that do not exceed 2.5 V for small photomultipliers (a few inches) or 7.5 V for larger ones.

The pulse shape in the program is modeled either as a double Gaussian shape , either as a Gamma function or it can be arbitrary. In any case the pulse shape is used in way so that the maximum of the pulse corresponds to the photoelectron creation time accounting also for the TTS.

The double Gaussian shape consists of two Gaussians with peak value 1.0 at the same time instant, but with different sigmas. The sigmas of the Gaussian functions are calculated by the rise and fall time of the pulse shape that are input by the user.

The form of the Gamma function used is:
$$F(t) = c\, t^a e^{-t/T} \quad ,$$
where c, a, T are constants that they are calculated from the rise and fall time that are input by the user, and the requirement that the maximum of the function to be 1.0.

If the user chooses to use an arbitrary pulse shape then he must provide a file with the binned values of this shape. The values are read from the class' constructor.

The class provides methods to:
- Return the single photoelectron waveform for a given time instant.
- Return the starting time of the waveform.
- Return the stopping time of the waveform.
- Return the charge (assuming 50Ω load) of the single photoelectron pulse. This is used to normalize any charge value so that in the end all charges are in single photoelectron units.
- Return the time difference between threshold cross and maximum. This is used to shift the timing of the threshold crosses later when from the waveform the PMT hits are estimated. The shift is such that for a single photoelectron the threshold crosses coincide with the photoelectron creation time taking into account the TTS.
- Return the timing of a threshold crossing for a waveform (or part of it) that is created from a single photoelectron. This timing depends on the charge of the pulse. This method is used for speed up since we do not have to search step by step in time for the threshold crossing when there is only one photoelectron creating the pulse.
- Return the ToT of a threshold crossing for a waveform (or part of it) that is created from a single photoelectron. This timing depends on the charge of the pulse and is used for speed up as is the previously mentioned method.

### 3.2.6 Class HitPE

This class is for providing access to the photoelectrons as they are produced by the HOURS_KM3Sim simulation. Each instance of this class is created for each photoelectron or for each group of synchronous photoelectrons on the same photomultiplier. The main objective of this class is to provide the amplitude of the single photoelectron pulse versus time. This amplitude is later used in class HitPMT in order of the full waveform on each PMT to be created.

From each line of the input file describing a photoelectron or a group of synchronous photoelectrons, the constructor resolves the line arguments and for each photoelectron of the group smears the time with the TTS and the charge with the charge resolution. Then in the case that the number of photoelectrons is large it groups the smeared photoelectrons in time intervals of 0.5 ns. This is done to reduce CPU usage for high energy events producing many synchronous photoelectrons.

The most important methods of this class provide:
- The amplitude of the waveform created by this group of photoelectrons versus time.
- The starting time of the same waveform. This is the pulse starting time of the first photoelectron of the group.
- The stopping time of the same waveform. This is the pulse stopping time of the last photoelectron of the group.

### 3.2.7 Class HitPMT

This class is to estimate and provide PMT hit information. A PMT hit is a threshold cross of the waveform of the photomultiplier and its properties are the timing of the threshold cross, the Time over Threshold (ToT) and the charge. A PMT waveform can have more than one PMT hit if the incident photoelectrons are timed far away enough from one to another. On the other hand, many photoelectrons that are close enough in time can produce one PMT hit. There is an instance of this class for every photomultiplier that has any kind of photoelectrons, signal or noise. This instance represents the waveform of this photomultiplier. All the photoelectrons (instances of the class HitPE) for a specific PMT are inserted to this class. After all HitPE instances are inserted the class is initialized. The initialization creates the PMT hits that could be more than one if the different HitPE instances represent photoelectrons that are sufficiently separated in time.

The estimation of the PMT hits is done by searching the PMT waveform for threshold crosses step by step in time. For events with many delayed photons the searching algorithm could spend a lot of time by searching in many microseconds where the waveform is zero. There is a method in this class that finds time intervals where there is no need to search for time over thresholds by searching step by step in time. These are the time intervals where the PMT waveform is zero and the time intervals where there is only one photoelectron. In the latter case there is a parametrization in class PulseShape that gives the information of threshold cross and ToT, so there is no need to search there for threshold crosses step by step in time.

In this class there is the option to deform the PMT waveform due to preamplifier effects. When the waveform exceeds a saturation limit at a given time, then the waveform is forced to take the value of saturation and the pulse there is lengthened in time. The increase of the pulse length depends on how much higher is the waveform compared to the saturation threshold. If there was not a charge loss then the ToT would become very large for many synchronous photoelectrons. However for a large number of photoelectrons measurements of KM3NeT front end electronics have shown that the maximum ToT reaches 200ns and then increases very slowly. To take into account that we don't increase the time step size of the pulse so that the charge under the waveform to be conserved. Instead, in every step we reduce the charge excess exponentially with coefficients that result in good agreement of ToT vs charge measurements observed experimentally. The constants that affect the deformation procedure are the saturation limit and the constants of the exponential reduction of the excess charge.

In this class there is also a simple parametrized function for the estimation of the charge from the ToT value. This function is valid for certain type of single photoelectron pulses and deformation procedures (pulse saturation algorithm) and it has been estimated from the KM3NeT optical modules.

### 3.2.8 Class HitOM

This is a class to provide Optical Module (OM) hit information. An OM hit is defined as a collection of PMT hits on the same optical module and within a predefined time interval. The OM hit except from the information inherited from the individual PMT hits (timing, charge and ToT) has also the information of directionality. This is the average direction of the active PMTs of the optical module. The weight in the average is the estimated charge of the PMT hit using its ToT value. An instance of this class is created when there are HitPMT instances that correspond to any of the PMTs of this Optical Module. These HitPMT instances are then loaded to the instance of the HitOM class and the PMT his are grouped in predefined time intervals. The timing of the OM hit corresponds to the timing of the first PMT hit in the group, the OM hit charge corresponds to the sum of all PMT hit charges and the OM hit ToT corresponds to the sum of the ToTs of all the PMT hits. The multiplicity of an OM hit is the number of PMTs having a PMT hit in the group, i.e. the number of PMT hits participating in this OM hit minus the number of PMT hits that are on the same PMT.

Taking into account the ToT for each PMT hit of an OM hit, as well as the multiplicity of the OM hit, the OM hit charge can be estimated with fair accuracy. This is because when the Optical Module consists of many small PMTs, a dense Cherenkov photon wave front do not activate only one PMT of the optical module but several ones. Put in a different way, the charge of the OM hit is correlated with the multiplicity of this hit. By parametrizing the deposited charge on an optical module versus the sum of the time over thresholds of the individual PMT hits, and this for every OM hit multiplicity, a good estimation can be made of the OM hit charge. Fig. 19 presents the accuracy of such an estimation, with the different lines representing different methods of the OM hit charge estimation.

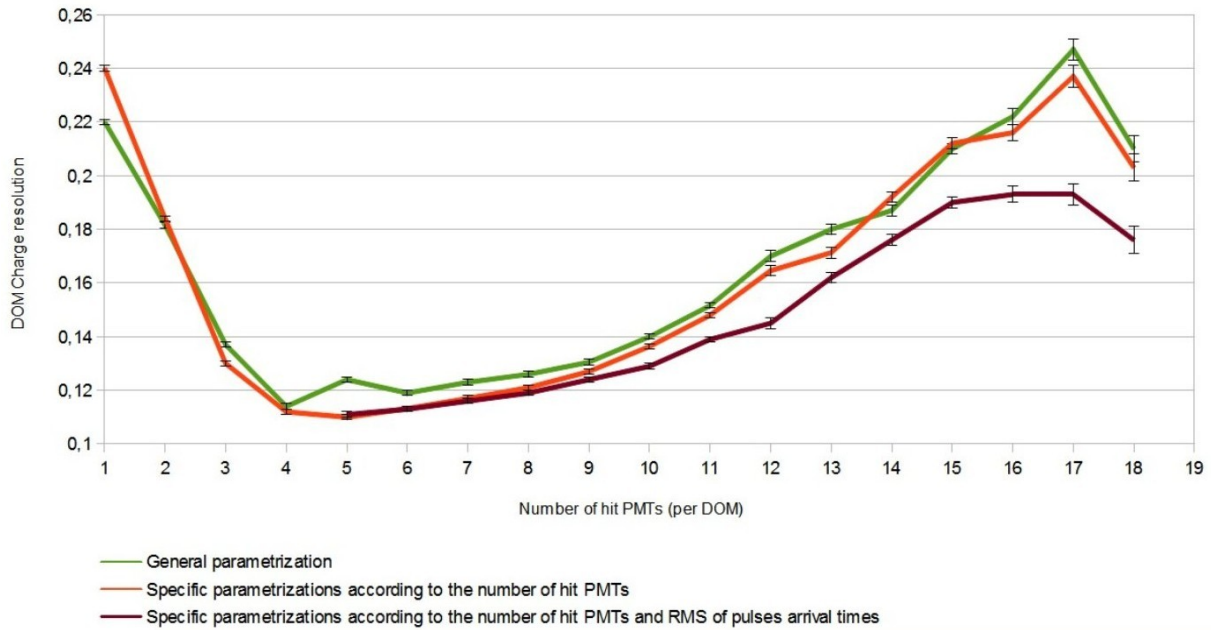

*Figure 19: The OM hit charge resolution versus the OM hit multiplicity for three kind of charge versus ToT parametrizations. The green line is for a general parametrization, where the multiplicity of the hit is not taken into account when building the parametrization. The red line is when a parametrization is build for every multiplicity. And the brown line is when a parametrization is build for every multiplicity and is also categorized according to the RMS of the arrival times of the individual PMT hits.*

### 3.3 Examples

There are two examples in the program distribution for running optical module simulation. The first example describes the detector response to optical noise photons from $^{40}$K decays, while the second example contains also photons from the secondaries of neutrino interactions.

The detector consists of 2070 Optical Modules on 115 vertical strings. The mean horizontal distance between strings is 90 meters. Each string carries 18 optical modules. The vertical distance between optical modules is 36 meters. The Optical Module consists of 31 3-inch photomultipliers.

For both examples the same input file with the photoelectron information created with HOURS_KM3Sim is used. This file contains the photoelectrons created by the secondaries of the neutrino interactions.

The optical noise background from $^{40}$K decays is used in both examples. The single and double multiplicity noise rate is input by the user, while the multiple (>2) noise rate is read from a file, as is explained in Section 3.2.2.

The Transit Time Spread (TTS) distribution is not considered Gaussian, but follows a binned distribution with bin values read from a file. The distribution is the one shown in Fig. 18.

The first example simulates the optical modules' response to $^{40}$K optical noise only. Although the input file is read, the photoelectrons are not used, and the only photoelectrons are those from $^{40}$K decays.

The information that the first example provides is the rate of various type of coincidences from $^{40}$K optical noise only. This information is read from the screen output and for each event contains:

- The number of active optical modules.
- The number of hits with single or higher multiplicity.
  - Single hits are those with no other hit in a 10 ns coincidence on the same optical module.
- The number of hits with double or higher multiplicity.
  - Double optical module hits are those with one other hit in a 10 ns coincidence on the same optical module.
- The number of hits with triple or higher multiplicity.
- The number of hits with quadruple or higher multiplicity.
- The number of hits with multiplicity 5 or more
- The number of hits with multiplicity 6 or more
- The number of hits with multiplicity 7 or more
- The number of pairs of neighboring optical modules with omhits causally connected and summed omhit multiplicities greater or equal to 2.
- The number of pairs of neighboring optical modules with omhits causally connected and summed omhit multiplicities greater or equal to 3.
- The number of pairs of neighboring optical modules with omhits causally connected and summed omhit multiplicities greater or equal to 4.
- The number of pairs of neighboring optical modules with omhits causally connected and summed omhit multiplicities greater or equal to 5.
- The number of pairs of neighboring optical modules with omhits causally connected and summed omhit multiplicities greater or equal to 6.
- The number of pairs of neighboring or next to neighboring optical modules with omhits causally connected and summed omhit multiplicities greater or equal to 2.
- The number of pairs of neighboring or next to neighboring optical modules with omhits causally connected and summed omhit multiplicities greater or equal to 3.
- The number of pairs of neighboring or next to neighboring optical modules with omhits causally connected and summed omhit multiplicities greater or equal to 4.

From these numbers we find the maximum and we apply it in the next example using the trigger mode 2.

For details see the doxygen documentation and the method TriggerEvent.

The second example is a run including noise and signal photoelectrons. The input file and the detector geometry is the same as in the first example. The triggering option used in this example uses the number of coincidences that have been found in the previous example, as it has been explained above.

# Appendix A
# Description of the EVT output.

In the following the EVT format is described for every step of the simulation in HOURS. This format is mainly derived from the ANTARES EVT format. The EVT format is a "tagged" format with each line's start having a tag describing the line. In the EVT format the output file has a header describing the input parameters of the simulation followed by several lines with each event's details.

The description is divided in three parts. In the first part the EVT format of the output of the generation part is described (HOURS_Gen), in the second part the output of the Cherenkov photon creation is described (HOUTS_KM3Sim), while in the third part the output of the Optical Module and front electronics simulation is described (HOURS_OmSim).

## A.1 EVT format for the Generation part of HOURS (HOURS_Gen)

In the generation part of HOURS the generated events follow two broad categories. The neutrino interaction events, where the PYTHIA generator is used, and atmospheric muon/neutrino generation events where the CORSIKA generator is used. In the second category there is also the option to produce atmospheric neutrino interactions in the vicinity of the detector. For these events both PYTHIA and CORSIKA are used. For both categories there are common tags in the output, but there are tags that appear only for the first kind of events and tags that appear only for the second kind of events. The following table describes the output header for both kind of event categories (brackets designate fixed values):

| tag | Parameter values | | | | |
|---|---|---|---|---|---|
| start_run: | Run number | | | | |
| spectrum: | Differential spectral index | | | | |
| physics: (only for CORSIKA events) | [CORSIKA] | [version] | [Several options with which CORSIKA is compiled] | | |
| physics: (only for PYTHIA events) | [PYTHIA] | [version] | | | |
| simul: | [HOURS_GEN] | [version] | | | |
| seed: | [HOURS_GEN] | "ranlux" random generator seed | | | |
| seed: (only for CORSIKA events) | [CORSIKA] | First CORSIKA random generator seed | Second CORSIKA random generator seed | | |
| seed: (only for PYTHIA events) | [PYTHIA] | Random number generator seed for PYTHIA | | | |
| model: | Generation mode | Type of neutrino or CORSIKA primary | Flag for selection of targets | Flag for Earth's shadowing | Flag for generation around detector |

| tag | | | | | |
|---|---|---|---|---|---|
| can: | Position of the bottom of the sea with respect to the detector center in m (negative real) | Can half height in meters | Can horizontal radius in meters | | |
| seabottom: | Sea depth in meters | | | | |
| genvol: | [0] | [0] | Depth of detector center | Radius of the event generation disk | Number of events simulated (including repetition number for CORSIKA events) |
| cut_nu: (for neutrino events) or cut_primary: (for CORSIKA events) | Minimum energy of simulated events in GeV | Maximum energy of simulated events in GeV | Minimum cosine of incident zenith angle for simulated events | Maximum cosine of incident zenith angle for simulated events | |
| cut_corsika: (only for CORSIKA events) | CORSIKA particle cut for hadrons | CORSIKA particle cut for muons | CORSIKA particle cut for electrons | CORSIKA particle cut for photons | |
| end_event: | | | | | |

For each event there are several lines describing its details. The following table describes these tags and the corresponding information written:

| tag | Parameter values | | | | | | | | | | |
|---|---|---|---|---|---|---|---|---|---|---|---|
| start_event: | Event number | [1] | | | | | | | | | |
| primary: (only for CORSIKA events) | [1] | [0] | [0] | [0] | x-coordinate of momentum vector | y-coordinate of momentum vector | z-coordinate of momentum vector | Energy of primary (GeV) | CORSIKA particle id code of primary | | |
| neutrino: (tag not appearing for CORSIKA events without atmospheric neutrino interactions | [1] | x,y,z coordinates of vertex in meters | x,y,z coordinates of neutrino direction vector | Neutrino energy in GeV | Starting event time in ns | X-BJORKEN variable | Inelasticity factor | Neutrino interaction channel (always -1 (DIS)) | Neutrino type according to PDG coding | CC or NC interaction flag (2=CC, 3=NC) | Target type according to PDG coding |

| track_bundle: (only for CORSIKA events) | Number of muons | Mean (weighted with energy) starting x-position of muons, in m | Mean (weighted with energy) starting y-position of muons, in m | Mean (weighted with energy) starting z-position of muons, in m | Mean (weighted with energy) starting x-direction of muons | Mean (weighted with energy) starting y-direction of muons | Mean (weighted with energy) starting z-direction of muons | Muon bundle energy in GeV | | | |
|---|---|---|---|---|---|---|---|---|---|---|---|
| track_in: | Particle's sequence number | Starting x-position of muon, in m | Starting y-position of muon, in m | Starting z-position of muon, in m | Starting x-direction of muon | Starting y-direction of muon | Starting z-direction of muon | Muon energy in GeV | Muon time in ns (relative to the fastest muon) | Particle type in GEANT 3 coding (6 for $\mu^-$ and 5 for $\mu^+$) | Particle type in PDG coding (13 for $\mu^-$ and -13 for $\mu^+$) |
| track_in: | 2nd muon's corresponding information |||||||||||
| | ... |||||||||||
| track_in: | last muon's corresponding information |||||||||||
| weights: | [0] ||| First event weight $(GeV\, m^2 sec\, sr)/year$ |||| Second event weight 1/year ||||
| probs_int: (only for CORSIKA runs when atmospheric neutrino interaction is enabled) | Interaction probability for all atmospheric neutrinos crossing the CAN |||  Interaction probability for the interacting atmospheric neutrino |||| | | | | |
| end_event: | |||||||||||

For CORSIKA events where the interaction of one of the shower's neutrinos is forced, the line with the tag "track_in" contains at the end one more flag that distinguishes atmospheric muons from the secondaries of the neutrino interaction.

**A.2 EVT format for the Cherenkov photon production part of HOURS (HOURS_KM3Sim)**

The information from the simulation of the Cherenkov photon production is added to the EVT file created in the previous step of the event generation. The information added includes information common for all events that is written in the file header, as well as information for each event. More detailed information can be found in the doxygen documentation. The following table describes the added information on the header of the EVT file:

| tag | Parameter values ||
|---|---|---|
| seed: | HOURS_KM3Sim | Random generator seed of HOURS_KM3Sim |

For every event that is simulated the information added is described in the following table:

| tag | Parameter values | | | | | | | | |
|---|---|---|---|---|---|---|---|---|---|
| hit: | ID number of hit | PMT ID number | Number of photoelectrons | Hit time | GEANT3 code of primary particle creating the hit | ID of primary track creating the hit | Number of photoelectrons before smearing for charge resolution | Hit time before smearing for TTS | ID of primary's particle physics process creating the hit |
| total_hits: | Number of hits | | | | | | | | |
| muonaddi_info: (When muon enter, is at center or leaves the detector) | Primary track ID | Position ID (-1 or 0 or 1) | x,y,z coordinates of position | x,y,z coordinated of unit direction vector | Energy at position (GeV) | Time at position | | | |
| muonaddi_info: (When muon stops inside detector) | Primary track ID | Position ID (2) | x,y,z coordinates of position | Time at position | | | | | |
| muon_decay: (When muon stops inside detector) | GEANT4 track id of secondary | primary muon track id | x,y,z position coordinates of secondary's creation point | x,y,z coordinated of unit direction vector of secondary | Energy of secondary (MeV) | Creation time of secondary | PDG particle code of secondary | | |
| muonenergy_info: (only when G4MYMUON_KEEPENERGY is defined at compilation and for single muons) | sequential number of line for this event | muon kinetic energy at a position (GeV) | muon kinetic energy at a position 10m after the previous one | ... | ... | ... | ... | ... | <<Maximum number of energy entries per line is 10>> |
| weight_sim: | Simulation weight | | | | | | | | |

**A.2 EVT format for the Optical Module and front end electronics simulation part of HOURS (HOURS_OmSim)**

The information from the simulation of the Optical Module and front end electronics is added to the EVT file created in the previous step of the Cherenkov photon production. The information added includes information common for all events that is written in the file header, as well as information for each event. More detailed information can be found in the doxygen documentation. The following table describes the added information on the header of the EVT file:

| tag | Parameter values | | |
|---|---|---|---|
| seed: | HOURS_OmSim | Random generator seed of HOURS_OmSim | |
| info_om: (appears only when the output hits are Optical Module hits) | ID number of first OM | x,y,z coordinates of OM position | x,y,z coordinates of OM direction |
| info_om: | ... | | |
| info_om: | ID number of last OM | x,y,z coordinates of OM position | x,y,z coordinates of OM direction |

For every event that is simulated the information added is described in the following table. Moreover the entries "hit:" and "total_hits:" of the previous simulation step are removed in order to reduce the output file size.

| tag | Parameter values | | | | | | | | | |
|---|---|---|---|---|---|---|---|---|---|---|
| hit_raw: (Appears only when output hits are PMT hits) | ID number of PMT hit | ID number of PMT | PMT hit charge in single pe units | PMT hit ToT | Estimated PMT hit charge in single pe units | PMT hit time | | | | |
| hit_rawOM: (Appears only when output hits are OM hits) | ID number of OM hit | ID number of OM | OM hit charge in single pe units | OM hit ToT | OM hit time | OM hit multiplicity | Zenith angle of the OM hit direction | Azimuth angle of the OM hit direction | The percentage of optical background in this OM hit | The expected OM hit time using the event direction |
| total_hits_raw: (Appears only when output hits are PMT hits) | Number of total PMT hits | | | | | | | | | |
| total_hits_rawOM: (Appears only when output hits are OM hits) | Number of total OM hits | | | | | | | | | |

**Appendix B**
**The algorithm for choosing emitted Cherenkov photons.**

As explained in the main text, during Cherenkov photon emission, only photons that are capable of reaching the detector optical modules are produced. For that purpose it is examined if a charged particle can emit light that can reach (unscattered) a sphere containing several OMs. This is accomplished by calculating the minimum and the maximum values of the photon emission zenith angles ($\theta_{min}$, $\theta_{max}$) in order to cross the sphere. For this calculation we apply a transformation where the particle direction lies along the positive z-axis and the center of the sphere in the (new) x-y plane (see Fig. 20).

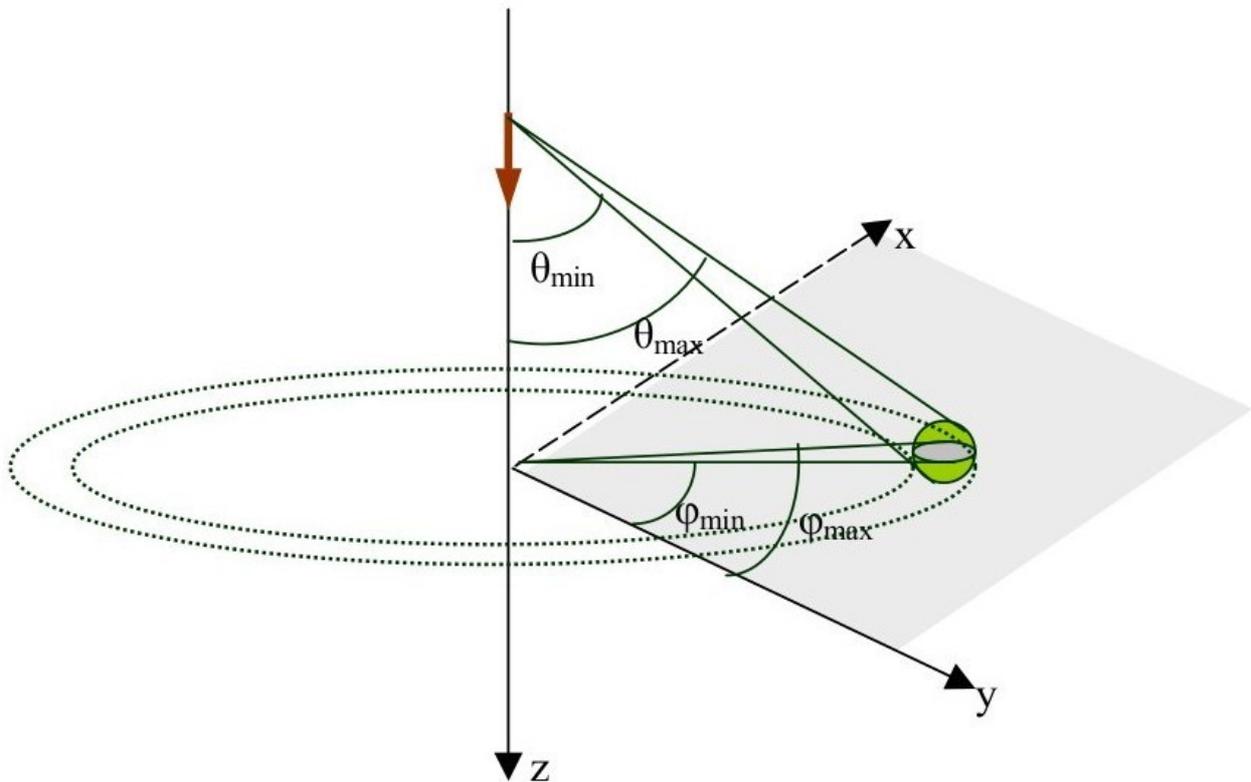

*Figure 20: In order to examine if a charged particle can emit Cherenkov photons that cross a sphere, we apply a transformation where the particle direction is along the positive z-axis and the sphere center is at the xy plane. Then we calculate the minimum and maximum angles ($\theta_{min}$, $\theta_{max}$, $\varphi_{min}$, $\varphi_{max}$) shown in the figure.*

By comparing these values with the corresponding extreme values of the Cherenkov emission angles (due to the variation of the refraction index in the wavelength range detectable by the photomultiplier) we can decide if we should produce or not the Cherenkov photons by the examined particle's step.

The searching algorithm takes into account the geometry of the detector and is divided in nested steps. In the first step we define a sphere that contains all of the detector, then we define 2 sub-spheres each containing half of the detector and so on. Each sphere is divided in two subspheres with the clustering method. The lowest level spheres are always the detector's OMs. The definition of the spheres with the clustering method is done in the class KM3Detector, where the detector geometry is also defined. The clustering method used ensures that each sphere is divided in two subspheres with almost equal number of OMs and the minimum possible radius.

In the previously described iterative check, the OMs that can be hit by a photon are marked keeping the information about the maximum and the minimum azimuth angles. Then for each Cherenkov photon we check if its azimuth angle is between the maximum and the minimum angle of the marked OMs and if so, we check if that photon can cross the OM, taking into account the photon's trajectory. If this is true then the photon is produced and the GEANT4 tracking algorithm is latched. A schematic view of the searching algorithm is shown in the Fig. 21.

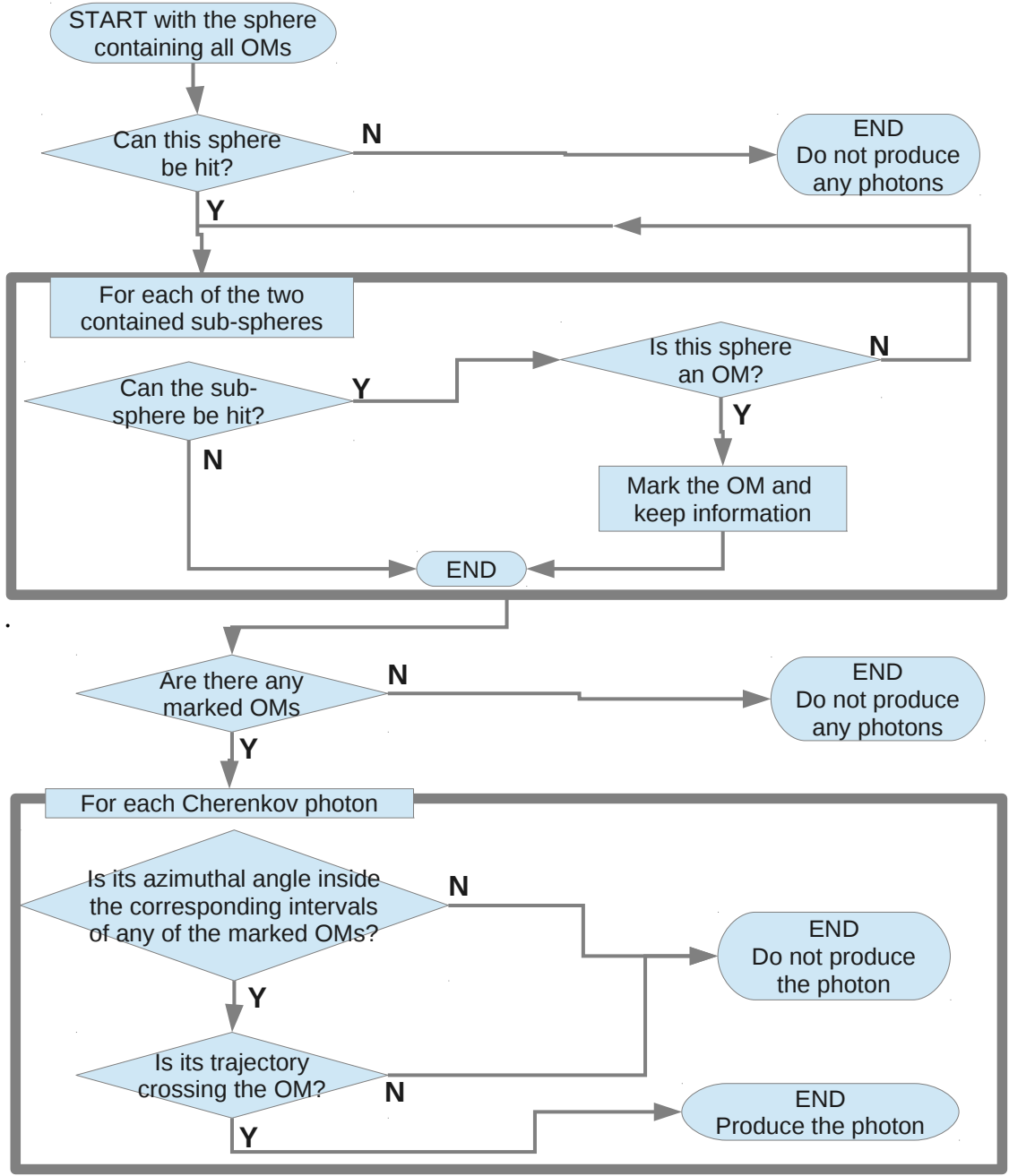

*Figure 21: A schematic view of the searching algorithm*